\documentclass[aps,prd,times,graphicx,amssymb,showpacs,groupedaddress,superscriptaddress,floatfix]{revtex4}
\usepackage{dcolumn}   % needed for some tables
\usepackage{bm}        % for math
\usepackage{multirow}
\usepackage{graphicx}
\usepackage{epstopdf}
\usepackage{amsmath}
\usepackage{amsfonts}
\usepackage{amssymb}
\usepackage{makeidx}
\usepackage{graphicx}
\usepackage{anysize}
\usepackage{hyperref}

\newcommand{\ba}{\begin{eqnarray}}
\newcommand{\ea}{\end{eqnarray}}
\newcommand{\be}{\begin{equation}}
\newcommand{\ee}{\end{equation}}

\newcommand{\pslashast}{\not{\!P^\ast}}

\newcommand{\nalpha}{\mbox{\boldmath $\alpha$}}

\newcommand{\np}{{\bf      p}}
\newcommand{\nq}{{\bf      q}}

\usepackage[dvips]{color}

\begin{document}

\title{Analysis of quasielastic $(e,e')$ electromagnetic responses and scaling for nuclear matter in the relativistic mean field model}
 \author{S.~Cruz-Barrios}
 \affiliation{Departamento de F\'{i}sica Aplicada I, Universidad de Sevilla, 41080 Sevilla, Spain}
 \author{J.A.~Caballero}
 \affiliation{Departamento de F\'{i}sica At\'omica, Molecular y Nuclear, Universidad de Sevilla, 41080 Sevilla, Spain}
\date{\today}

\begin{abstract}
A detailed study of the electromagnetic responses for quasielastic $(e,e')$ reactions from nuclear matter is presented within the framework of a relativistic model including momentum dependent scalar and vector mean field potentials in both the initial and final nucleon wave function states. The effects ascribed to the use of energy-dependent potentials are carefully analyzed by comparing their predictions with the responses obtained in the case of constant, i.e., energy-independent, potentials, as well as with the plane wave limit in the final state. The study is extended to the scaling phenomenon. Results are provided for the scaling functions corresponding to different nuclear systems evaluated at several kinematics ranging from intermediate to high values of the momentum transfer. Emphasis is placed on scaling of the first and second kinds and the role played by the relativistic scalar and vector mean field potentials.
\end{abstract}

% \pacs{13.15.+g, 25.30.Pt}

\maketitle

\section{Introduction}

In recent years physicists have devoted a great effort to the investigation of neutrino properties through the analysis of neutrino oscillation experiments~\cite{Alvarez-Ruso:2017oui,Katori:2016yel}. The main objective is to measure the leptonic CP violation phase, assessing the neutrino mass hierarchy, and improve our present knowledge on the oscillation mixing angles. As most of the detectors used in neutrino oscillation experiments consist typically of complex nuclear systems, i.e., carbon, argon, water or mineral oil, it is mandatory to describe accurately how neutrinos interact with nuclei at very different kinematical regimes. Not only the quasielastic (QE) region should be carefully evaluated but also the $\Delta$ and higher nucleon excitations, two-particle two-hole (2p-2h) effects and even deep inelastic scattering (DIS). Also the region of very low momentum transfer, in which the impulse approximation is highly questionable, can have a significant impact in the analysis of results. As known, this makes an important difference from the case of electron beams. Not only the interaction, electromagnetic versus weak, is different but also the kinematics of the processes. Whereas the electron beam energy is known with accuracy, so one has control on the main reaction channel involved in the process, in the case of neutrinos the energy range covered by the beams can be extended from hundreds of MeV to several GeV. Hence in neutrino experiments nuclear effects corresponding to different reaction mechanisms can play an important role and should be under control in order to provide a coherent analysis and interpretation of data. Furthermore, the high energies involved require relativity as an essential ingredient. 

As a general rule any model aiming to describe neutrino-nucleus interaction processes should be first tested against electron scattering data. Only after the nuclear model is validated using high quality electron scattering data in the relevant energy domain, one can think of extending it to nuclear weak processes. This has been often stressed in previous works where careful comparisons between theoretical predictions based on different nuclear models and electron scattering data collected in the quasielastic scattering (QES) domain have been presented~\cite{Amaro:2010sd,Megias:2016lke,Megias:2017cuh,Bodek:2016abf,Bodek:2014jxa,Bodek:2014pka,Kim:2007yj,Kim:2003wy,Giusti:2013gsa,Meucci:2013pua,Rocco:2015cil,Benhar:2015xga,Benhar:2010nx,Mosel:2018qmv}. 
Although these studies have been very successful, the complexity of the quantum many-body problem is still far from being resolved. 
In a series of previous works~\cite{Caballero:2006wi,Maieron:2001it,Megias:2016fjk,Gonzalez-Jimenez:2014eqa,Amaro:2004bs,Amaro:2010sd,Amaro:2011aa,Caballero:2005sj} we have analyzed in detail the properties of scaling and superscaling in quasielastic inclusive $(e,e')$ data. The importance of this phenomenon to test the validity of any nuclear model aiming to describe electron scattering reactions, and its impact when applied to neutrino-nucleus interactions, has been clearly proved. The SuperScaling Approach (SuSA), and its improved version, denoted as SuSAv2~\cite{Gonzalez-Jimenez:2014eqa,Megias:2016lke,Megias:2016fjk}, are based on the existence of "universal" scaling functions to be valid not only for electrons but also for neutrinos. The SuSA model is entirely based on the phenomenology, making use of a unique, universal, scaling function extracted from the analysis of the longitudinal electron scattering data. This phenomenological function is taken to be the same for the transverse electromagnetic channel as well as for all the neutrino responses. On the contrary, the SuSAv2 model incorporates relativistic mean field (RMF) effects in the longitudinal and transverse nuclear responses, as well as in the isovector and isoscalar channels that is of great importance in order to describe charged-current (CC) neutrino reactions that are purely isovector~\cite{Caballero:2007tz}.

The origin of the SuSAv2 approach is based on the capability of the RMF to describe properly the scaling behavior of the electron scattering data. As shown in previous works~\cite{Caballero:2007tz,Caballero:2005sj,Caballero:2006wi}, RMF is one of the few microscopic models capable of reproducing the asymmetric shape of the phenomenological scaling function with a long tail extended to high values of the transfer energy. Moreover, RMF produces an enhancement in the transverse scaling function, a genuine dynamical relativistic effect linked to the lower components in the wave functions, that is supported by the analysis of data. The RMF framework to finite nuclei has proven to successfully reproduce the scaling behavior shown by data at low to intermediate $q$ values. However, the model clearly fails at higher momentum transfers where Final State Interactions (FSI) are expected to be weaker. This is so because of the use of very strong energy-independent scalar and vector potentials in the final state that lead to too much asymmetry in the scaling functions apart from shifting the QE peak to very high transfer energies, in clear disagreement with data. To remedy this shortfall of the RMF model, the SuSAv2 incorporates both the pure RMF scaling functions at low-to-intermediate $q$ values and the Relativistic Plane Wave Impulse Approximation (RPWIA) ones at higher $q$ by using a $q$-dependent blending function that smoothly connects both regimes. Although SuSAv2, once 2p-2h meson exchange currents (MEC) are also incorporated, works properly in most of the cases providing an excellent agreement with $(e,e')$ data in the whole energy range, it would be also highly desirable to provide a unified model that could address consistently both regimes, i.e., from low-to-intermediate up to very high $q$-values. Work along this line is presently in progress by describing the outgoing nucleon wave function as a solution in the continuum of the Dirac equation with energy-dependent potentials~\cite{Raul:2019}. 

In this work our approach to the problem is however different. We place our focus on the failure of the pure RMF model in describing the electromagnetic responses at high $q$-values. As already mentioned, the use of energy-independent scalar and vector potentials in the final nucleon states leads to FSI effects that do not diminish as the momentum transfer reaches very high values. This is not yet fully understood, and it shows a clear deficiency of the model. In this work we present a systematic analysis of the problem, but making use of a simplified description of the electromagnetic responses based on the relativistic mean field model applied to nuclear matter. This strategy, that follows the previous studies in \cite{Horowitz:1993nb,Kim:1994nr}, allows us to provide analytical expressions for the tensors and response functions in most of the cases and show up in a very transparent way the role played by the relativistic potentials in the nuclear electromagnetic responses. A variety of situations is investigated by considering different options for the potentials, namely, dependence/independence on the energy, and use of the same and/or different potentials in the initial/final states. Furthermore, a very detailed study of the scaling properties is also provided. Scaling functions are evaluated with the different model descriptions with particular emphasis on how the responses fulfill scaling of the first and second kinds and its connection with the specific role played by the relativistic scalar and/or vector potentials. 

The paper is organized as follows: In Sec. II we introduce the basic formalism and present the general expressions involved in describing the electromagnetic $(e,e')$ response functions within the framework of the relativistic mean field model applied to nuclear matter. The most general expression of the polarization tensor, valid for different descriptions of the initial and final nucleon states, with its corresponding components to be applicable to the calculation of the longitudinal and transverse response functions is shown and discussed in detail. A study focused on scaling and superscaling properties is also provided. In Sec. III we present our results for the electromagnetic longitudinal and transverse response functions corresponding to very different kinematical situations and nuclear targets. A detailed discussion of the results obtained for the scaling functions is also shown. Finally, in Sec. IV we summarize our conclusions. 

\section{General formalism}

The general formalism for $(e,e')$ reactions has been presented in previous works~\cite{Donnelly:1985ry,Donnelly:1998xg}. Assuming Plane Wave Born Approximation (PWBA), i.e., one virtual photon exchanged and leptons described as free particles, the QE differential cross section can be expressed in terms of two nuclear response functions:
\be
\frac{d\sigma}{d\varepsilon' d\Omega'}=\sigma_M\left[v_LR^L(q,\omega)+v_TR^T(q,\omega)\right] \, , \label{eq1}
\ee
where $\varepsilon'$ ($\Omega'$) is the scattered electron energy (solid angle) and $\sigma_M$ represents the Mott cross section. The kinematic factors $v_L$ and $v_T$ come solely from the leptonic tensor and their explicit expressions can be found in \cite{Donnelly:1985ry,Amaro:2002mj}. Finally, $R^L$ and $R^T$ are the electromagnetic response functions that contain the whole dependence on the nuclear vertex coupling and depend on the momentum ($q$) and energy ($\omega$) transferred in the process. The notation $L$ ($T$) refers to longitudinal (transverse) components with respect to the direction of the momentum transfer $\nq$.

The response functions can be evaluated by taking the appropriate components of the polarization propagator $\Pi^{\mu\nu}$ (also referred to as the current-current correlation function)~\cite{Kim:1994nr,Horowitz:1993nb,Amaro:2002mj}
\ba
R^{L} \,&=&\, -\frac{2}{\pi \rho}\textrm{Im}\left\{Z \Pi_{p}^{00} +N \Pi_{n}^{00}\right\}  \label {RL}\\
R^{T} \,&=& \,-\frac{2}{\pi \rho}\textrm{Im} \left\{Z \left(\Pi_{p}^{11}+\Pi_{p}^{22}\right)+N\left( \Pi_{n}^{11}+\Pi_{n}^{22}\right)\right\} \, ,
\label{RT}
\ea
where $\Pi^{\mu\nu}_{p(n)}$ refers to the polarization propagator for protons (neutrons), and $Z$ ($N$) represents the number of protons (neutrons) in the nuclear target. We use a coordinate system with the $z(3)$-axis in the direction of $\nq$. Note that if gauge invariance is fulfilled, the $3$ component in the current is connected with the $0$ (time) component. This explains that the longitudinal response is simply given by the time component of the polarization propagator. On the contrary, 1 and 2 refer to the transverse components. 

The nuclear responses in (\ref{RL}) and (\ref{RT}) are evaluated in a local density approximation from nuclear matter with the density given by $\rho=2k_F^3/(3\pi^2)$, and $k_F$ the Fermi momentum whose specific value for the different nuclear systems investigated in this work will be given later. Since closure can be used to carry out the sum over final states, the polarization propagator can be expressed in terms of the full propagator, $\hat{G}$, of the nuclear many-body system, 
\be
\Pi_{p(n)}^{\mu\nu}(q,\omega)=  -i\int\frac{d^4P}{(4 \pi)^{4}} Tr[\hat{G}(P+Q) 
\Gamma_{p(n)}^{\mu}\hat{G}(P) \Gamma_{p(n)}^{\nu}] \, ,
\ee
where $\hat{G}(P)$ represents the Green function for a nucleon propagator and $\Gamma_{p(n)}^\mu$ is the
electromagnetic vertex corresponding to protons (neutrons). Here we use the relativistic free nucleon expression~\cite{DeForest:1983ahx} 
\be
\Gamma_{p(n)}^{\mu} = F^{p(n)}_{1}\gamma^{\mu} + 
F^{p(n)}_{2}\frac{i\sigma^{\mu\nu}Q_{\nu}}{2M}
\ee  
with $F_1^{p(n)}$ and $F_2^{p(n)}$ the Pauli and Dirac proton (neutron) form factors, respectively, that depend only on the transferred four-momentum, $Q^2=\omega^2-q^2$. In this paper the proton and neutron form factors used correspond to the well-known Galster parametrization~\cite{Galster:1971kv}. The term $M$ represents the mass of the nucleon.
The corresponding Dirac equation
%nucleon self-energy solution of the Dirac equation 
is written as
\be
\left[\nalpha\cdot \np +\beta\left(M+S(\np)\right)+V(\np)\right]U(\np)=E_\np U(\np) \, ,
\ee
%\be
%\Sigma=\gamma_0V(\np) +S(\np) \, ,
%\ee
where $\nalpha$ and $\beta$ are the Dirac matrices, and 
$S$ and $V$ represent the relativistic scalar and vector potentials that may include dependence on the energy-momentum. In our case we take $S$ ($V$) from a Dirac optical potential fit at the self-consistent energy $E_\np$,
\be
E_{\np} \,= \, \sqrt{ \np^{2}+[M+S(\np)]^{2}}+V(\np) \, .
\ee
Thus, the Green function evaluated in a mean field approximation and corresponding to a value of the Fermi momentum $k_{F}$ is given by
\be
\hat{G}(P) = (\pslashast + M^{*}_{\np}) \biggl[
\frac{1}{P^{*^{2}}-M^{*^{2}}_{p}+i\epsilon} + \frac{i \pi}{E^{*}_{\np}}
\delta(p_0\, - \, E_{\np})\Theta(k_{F}-|\np|) \biggr] \, ,
\ee
where we use the standard Feynman notation, i.e., $\pslashast\equiv P^{\ast\mu}\gamma_\mu$, and have introduced the effective nucleon four-momentum:
\be
P^{\ast\mu}=(p^{0}-V(\np) \,  , \,  \np ) \; = \; 
      (E_{\np} -V(\np)\, , \, \np ) =(E^\ast_\np\, , \, \np) \, \label{Pstarmu}
\ee
with the energy $E^\ast_\np = \, \sqrt{\np^{2} \, + \, M^{*^{2}}_{\np}}$ expressed in terms of the effective mass $M^\ast_\np$, i.e., the nucleon mass modified by the presence of the scalar potential, $M^\ast_\np \equiv M+S(\np)$.
 
Finally, integrating over $p_0$ the imaginary part of the polarization propagator becomes
\be 
\textrm{Im}\, \Pi^{\mu\nu}= -\int \frac{\np^2 d|\np|d\cos\theta}
{4 \pi E^{*}_{\np_{i}}E^{*}_{\np_{f}} }\Theta(|\np_{f}|-k_{F})
\Theta(k_{F}-|\np|)
\delta(E_{\np_{f}}- E_{\np_{i}} - q_{0}) T^{\mu\nu}(p,q,\omega) \, ,
\ee 
where
$P^{\mu}_{f} = (P_{i} + Q)^{\mu}$ and $\theta$ is the angle between $\np_{i}$ and $\nq$. After a laborious algebra involving the trace of Dirac matrices, the final expression for the tensor $T^{\mu\nu}$ results
\ba
T^{\mu\nu}  &= &
	g^{\mu\nu} \biggl\{ 
          \Bigl[ M^{*}_{i}M^{*}_{f} - (P^\ast_{i} \cdot P^\ast_{f}) \Bigr]F^{2}_{1}
+  \Bigl[ 2M^{*}_{i}(P^{*}_{f} \cdot Q)-2M^{*}_{f}(P^{*}_{i} \cdot Q) \Bigr]
                       \frac{F_{1}F_{2}}{2M}  
                      \nonumber \\
     &+&  \Bigl[ \bigl(M^{*}_{f}M^{*}_{i} + P^{*}_{i} \cdot P^{*}_{f} \bigr)Q^{2} -
                2\bigl(P^{*}_{i} \cdot Q \bigr) \bigl( P^{*}_{f} \cdot Q \bigl) \Bigr]
                    \frac{F^{2}_{2}}{4M^{2}}   \biggl \} 
                    \nonumber \\
     & + &  \Bigl(P^{*^{\mu}}_{f}P^{*^{\nu}}_{i} + P^{*^{\nu}}_{f}P^{*^{\mu}}_{i} \Bigr)
             \Bigl[F^{2}_{1} - \frac{Q^2}{4M^{2}}F^{2}_{2} \Bigr] 
		-Q^{\mu}Q^{\nu} \Bigl[M^{*}_{i}M^{*}_{f} + 
                           P^{*}_{i} \cdot P^{*}_{f} \Bigr] \frac{F^{2}_{2}}{4M^{2}}
\nonumber \\   
	&+& \Bigl(Q^{\mu}P^{*^{\nu}}_{i} + Q^{\nu}P^{*^{\mu}}_{i} \Bigr)
            \Bigr[M^{*}_{f}\frac{F_{1}F_{2}}{2M} + \bigl(P^{*}_{f} \cdot Q \bigr)
                    \frac{F^{2}_{2}}{4M^{2}} \Bigr] 
                    \nonumber \\
    & + &    \Bigl(Q^{\mu}P^{*^{\nu}}_{f} + Q^{\nu}P^{*^{\nu}}_{f} \Bigr)
             \Bigl[-M^{*}_{i }\frac{F_{1}F_{2}}{2M} + \bigl(P^{*}_{i} \cdot Q \bigr)
                  \frac{F^{2}_{2}}{4M^{2}} \Bigr]    \, .
\ea
In order to evaluate the specific components of the polarization tensor corresponding to the longitudinal and transverse response, we distinguish between the variables, $P^{\ast\mu}=(E_\np-V(\np)\, , \, \np)$ and $P^\mu=(E_\np\, ,\, \np)$ that apply to both the initial ($\np_i$) and final ($\np_f$) states. Taking into account energy-momentum conservation, i.e., $P^\mu_f=(P_i+Q)^\mu$, we can introduce an effective transfer four-momentum given as
\be
Q^{\ast\mu}=(P^\ast_f-P^\ast_i)^\mu=Q^\mu-\Delta V^\mu \, ,
\ee
where we have introduced the notation $\Delta V^\mu \equiv (V(\np_f)-V(\np_i)\, ,\, 0)$. Then, note that $Q^{\ast\mu}$ and $Q^\mu$ only differ in the time component, that is, in the energy transfer that satisfies: $\omega^\ast=\omega - \Delta V$ with $\Delta V\equiv V(\np_f)-V(\np_i)$. Finally, introducing also the effective masses, $M^\ast_{i,f}$, from the corresponding energies, namely, 
$E^\ast_{\np_{i,f}}=\sqrt{M^{\ast 2}_{i,f}+\np_{i,f}^2}$, the following results emerge:
 \ba
P^{*}_{f} \cdot P^{*}_{i} & = &  \frac{1}{2} 
     \Bigl( M^{*^{2}}_{f} + M^{*^{2}}_{i} - Q^{*^{2}} \Bigr) \nonumber \\
 P^{*}_{f} \cdot Q  & = & \frac{1}{2} 
        \Bigl( M^{*^{2}}_{f} - M^{*^{2}}_{i} + Q^{*^{2}} \Bigr) + E^{*}_{\np_f} \Delta V 
  \nonumber \\
P^{*}_{i} \cdot Q  & = & \frac{1}{2} 
          \Bigl(M^{*^{2}}_{f} - M^{*^{2}}_{i} -Q^{*^{2}} \Bigr) + E^{*}_{\np_i} \Delta V \, .
\ea
In what follows we evaluate explicitly the specific components of the tensor that enter in the longitudinal and transverse response functions.
 
%%%%%%%%%%%%%%%%%%%%%%%%%%%%%%%%%%%%%%%%%%%%%%%%%%%%%%%%%%%%%%%%%%%%%%%%%%%%%

%\subsection{Longitudinal channel}
As shown in (\ref{RL}) the longitudinal response is given by the time-time component of the tensor, $T^{00}$. After some algebra we can write
\ba
T^{00} & = & \frac{1}{2} \bigg[Q^{*^{2}} - \Big(M^{*}_{f} -M^{*}_{i}\Big)^{2}
                      \bigg] \bigg[F_{1} \, + \, 
                    \Big(M^{*}_{f}+M^{*}_{i}\Big)\frac{F_{2} }{2M}\bigg]^{2}
                \nonumber \\
         &  &   + 2E^{*}_{\np_i}\Big(E^{*}_{\np_i}+\omega^{*}\Big)\bigg[F^{2}_{1} -
                      \frac{Q^{*^{2}}}{4M^{2}}F^{2}_{2} \bigg]
                      \nonumber \\
      &   &     + \omega^{*}E^{*}_{\np_i} \bigg[ \frac{\Big(M^{*}_{f} -M^{*}_{i} \Big)}{M}
                   F_{1}F_{2} \,+ \, 2\frac{\Big(M^{*^{2}}_{f}-M^{*^{2}}_{i} \Big)}{4M^{2}}
                      F^{2}_{2} \bigg] 
                      \nonumber \\
       &  &     + \frac{\omega^{*^{2}}}{2} \bigg[  - \frac{Q^{*^{2}}}{4M^{2}}F^{2}_{2}
                      - \frac{ \Big(M^{*}_{i} + M^{*}_{f} \Big)^{2}}{4M^{2}}F^{2}_{2}
                      + 2\frac{\Big(M^{*^{2}}_{f} - M^{*^{2}}_{i}\Big)}{4M^{2}}F^{2}_{2}
                     -2\frac{M^{*}_{i}}{M}F_{1}F_{2}\bigg]  \, .
                     %\nonumber \\
\ea  
To make contact with some previous works~\cite{Amaro:2002mj,Amaro:2004bs,Maieron:2001it} we introduce a set of dimensionless variables:
\ba
&&
\kappa \equiv \frac{q}{2M}; \qquad  \lambda \equiv \frac{\omega}{2M}; \qquad  
            \eta_i \equiv \frac{p_i}{M}; \qquad  \tau = \kappa^{2} - \lambda^{2}
                  \nonumber \\
         &&         s_{m_{i}} \equiv \frac{S(\np_i)}{M};  \,\,
                     \Delta v \equiv \frac{\Delta V}{2M} ; \,\,
            \lambda^{*} \equiv \frac{\omega^{*}}{2M}=\lambda-\Delta v;  \,\,  \tau^{*} = \kappa^{2} - \lambda^{*^{2}}                  
                     \nonumber \\
                     &&
 	\epsilon^{*}_{i} \equiv \frac{E^\ast_{\np_i}}{M}= \sqrt{\eta_i^{2} + (1+s_{m_{i}})^{2}}; \,\,
           \Delta m^{*^{2}} \equiv  \frac{1}{4} 
        \bigg[ \Big(\frac{M^{*}_{f}}{M}\Big)^{2}-\Big(\frac{M^{*}_{i}}{M}\Big)^{2} \bigg];
                             \,\,
           \rho^{*} = \Big( 1+ \frac{\Delta m^{*^{2}}}{\tau^{*}} \Big)   \, .   \label{adimension}
\ea 
The longitudinal tensor results:
\ba
T^{00} & = & 2M^{2} \Bigg \{ - \bigg(\tau^{*} + \frac{1}{4} 
                   \Big(s_{m_{f}}-s_{m_{i}} \Big)^{2} \bigg) \bigg[F_{1} + 
                         \Big(1 + \frac{s_{m_{f}}+s_{m_{i}}}{2} \Big)F_{2} \bigg]^{2} 
                        \nonumber \\
     &     &    +  \bigg(\epsilon^{*^{2}}_{i} + 2 \lambda^{*} \epsilon^{*}_{i} \bigg)
                    \bigg[F^{2}_{1} + \tau^{*}F^{2}_{2} \bigg]
	 + 2 \lambda^{*}\epsilon^{*}_{i} \bigg[ \frac{1}{2} 
                   \Big(s_{m_{f}} -s_{m_{i}} \Big)F_{1}F_{2} + \Delta m^{*^{2}}F^{2}_{2} \bigg]
                           \nonumber \\
    &      &    + \lambda^{*^{2}} \bigg[ \Big(\tau^{*}+ 2 \Delta m^{*^{2}} - 
                         \big(1 + \frac{s_{m_{f}}+s_{m_{i}}}{2} \big)^{2} \Big)F^{2}_{2}
                        -2 \Big( 1+ s_{m_{i}} \Big) F_{1}F_{2} \bigg] \Bigg \} \, . \label{T00}
\ea
Note that the whole dependence on the self-energy vector potential $V(\np)$ only enters through the 
$\tau^{*}$ and $\lambda^{*} $ terms.

%\subsection{Transverse channel}
For the transverse channel (\ref{RT}) we have to evaluate the two pure transverse components in the tensor. After some algebra one can finally write,
\ba 
T^{11} +T^{22}& = & - \Bigg\{ \bigg[Q^{*^{2}}- \Big(M^{*}_{f} - M^{*}_{i} \Big)^{2} \bigg]
                        \bigg[F_{1} + \Big( M^{*}_{f} + M^{*}_{i} \Big)\frac{F_{2}}{2M} \bigg]^{2}
                         \nonumber \\
               &   &         + 2 \Delta V \bigg[ \frac{\Big(M^{*}_{i}-M^{*}_{f} \Big)}{M}E^{*}_{i} 
                                   + \frac{M^{*}_{i}}{M} \omega^{*} \bigg]F_{1}F_{2}
                                 \nonumber \\
          &   &         + \bigg[ \Delta V^{2} 
                                \Big[ \frac{\big(M^{*}_{f}+M^{*}_ {i}\big)^{2}}{4M^{2}}
                                    -\frac{Q^{*^{2}}}{4M^{2}}
                                    -4\frac{E^{*}_{i}}{4M^{2}}\big(E^{*}_{i}+\omega^{*}\big)\Big]
                                        \nonumber \\
           &    &                    + 2 \Delta V 
                                   \Big[\omega^{*}\frac{\big(M^{*}_{i}+M^{*}_{f} \big)^{2}}
                                         {4M^{2}}
                                        -2\frac{\big(M^{*^{2}}_{f}-M^{*^{2}}_{i}\big)}{4M^{2}} 
                                         \big(2E^{*}_{i} + \omega^{*}\big) \Big] \bigg]F^{2}_{2}
                                       \nonumber \\
            &     &         +  2|\np|^{2} \Big( 1 - \cos^{2} \theta \Big) 
                                    \bigg[F^{2}_{1} - \frac{Q^{2}}{4M^{2}}F^{2}_{2} \bigg] 
                                         \Bigg \} \, .
\ea
In terms of the dimensionless variables introduced in (\ref{adimension})
\ba
&&	T^{11}+T^{22} =   4M^{2} \Bigg \{ \bigg[\tau^{*} +\frac{\Big(s_{m_{f}}-s_{m_{i}}\Big)^{2}}{4}\bigg]
                                 \bigg[F_{1} + \Big(1 + \frac{s_{m_{f}}+s_{m_{i}}}{2} \Big)F_{2} 
                                           \bigg]^{2} 
                                                 \nonumber \\
          &    &              +\Delta v \bigg[\Big(s_{m_{f}}-s_{m_{i}} \Big) \epsilon^{*}_{i}
                                        -2 \Big(1+s_{m_{i}} \big)\lambda^{*} \bigg]F_{1}F_{2}
                                                     \nonumber\\
            &    &                + \bigg[ \Delta v^{2}
                                           \Big( \epsilon^{*^{2}}_{i} 
                                               + 2\epsilon^{*}_{i} \lambda^{*} -\tau^{*}
                                               -\big( 1+ \frac{s_{m_{f}}+s_{m_{i}}}{2} \big)^{2} 
                                                \Big)
                                                       \nonumber \\
           &     &                         + 2\Delta v \Big( 
                                        \Delta m^{*^{2}} \big(\epsilon^{*}_{i}+\lambda^{*} \big)
                                  - \lambda^{*} \big(1 + \frac{s_{m_{f}}+ s_{m_{i}}}{2} \big)^{2} \Big)
                                                \bigg]F^{2}_{2}
                                          \nonumber \\ 
           &     &            + \frac{1}{2}\bigg[ \frac{\tau^{*}}{\kappa^{2}}  
                                                  \Big(\epsilon^{*^{2}}_{i} 
                                          + 2\epsilon^{*}_{i}\lambda^{*}\rho^{*} +\lambda^{*^{2}}
                                           -\Delta m^{*^{2}} \big(1+ \rho^{*} \big) \Big)
                                           -\Big( \tau^{*} + \big(1+s_{m_{i}} \big)^{2} \Big) 
                                              \bigg] \bigg[F^{2}_{1} + \tau F^{2}_{2} \bigg]
                               \Bigg \} \, .
%                             \nonumber \\
\ea
From the above expressions for the two channels the polarization tensor can be evaluated by solving the integrals numerically. Results for the response functions are presented in next section. Here we assume the phenomenological scalar and vector potentials, $S(\np), \, V(\np)$, to be fitted to polynomials in $\np$. Their explicit expressions are given in appendix A,  and they are consistent with the analysis of elastic electron scattering data at high energies. 

%\subsection{Off Shell effects}
In what follows we restrict our attention to some simplified situations in which the integrals can be solved analytically, so explicit expressions for the polarization tensor, likewise for the nuclear response functions, can be obtained. As already mentioned, the whole dependence on the vector potential $V$ enters through the terms $\tau^\ast$ and $\lambda^\ast$. In the particular case of assuming the vector potentials in the initial and final states to be almost equal, namely, $V(\np_i)\simeq V(\np_f)$, the term $\Delta V$ tends to zero, and the effective energy transfer $\omega^\ast$ coincides with $\omega$. Thus, in the limit case of vector potentials being equal in the initial and final states, the tensor (likewise the polarization propagator) does not depend on $V$ and the only dependence on the scalar potential $S$ enters through the effective masses: $M^\ast_{i,f}$. In terms of the dimensionless variables, the explicit expression for the tensor in the longitudinal channel is analogous to (\ref{T00}) but changing $\tau^*$ and $\lambda^*$ by $\tau$ and $\lambda$, respectively. In the case of the transverse channel, the final result is given in the form:
\ba 
T^{11}+T^{22} & = & 2 M^{2} \Bigg \{ 2 \bigg[\tau + \frac{ \Big(s_{m_{f}}- s_{m_{i}} \Big)^{2}}{4}
                              \bigg]
                                        \bigg[F_{1} + \Big( 1+ \frac{s_{m_{f}}+s_{m_{i}}}{2} \Big)
                                                    F_{2} \bigg]^{2}
                             \nonumber \\
                             &   & + \bigg[\frac{\tau}{\kappa^{2}} \Big( \epsilon^{*^{2}}_{i} + 
                              2\epsilon^{*}_{i}\lambda\rho^{*} + \lambda^{2} - \Delta m^{*^{2}} 
                        \big(1 + \rho^{*} \big) \Big) - \Big( \tau+ \big(1+s_{m_{i}} \big)^{2}
                                                      \Big) \bigg]  
                                   \bigg[F_{1}^{2}+ \tau F_{2}^{2} \bigg] \Bigg \} \, .
%                                \nonumber \\
\ea                           
Finally in the simple case of constant scalar and vector potentials the integrals in the polarization propagator 
can be solved analytically.
The final expressions for the components of $\Pi^{\mu\nu}$ (assuming equal constant potentials in the initial and 
final states) are:
\ba 
-\textrm{Im}\, \Pi^{00} & = & \frac{M^{2}}{4\pi \kappa} \Big( \epsilon^{*}_{F}- \Gamma \Big)
                                 \Theta \Big( \epsilon^{*}_{F} - \Gamma \Big) 
                       \Bigg \{ -\kappa^{2} \bigg[F_{1} + \Big( 1+ s_{m} \Big)F_{2} \bigg]^{2}
                                  \nonumber \\
            &   & + \frac{\kappa^{2}}{ \tau} 
                       \bigg[ \Delta + \tau + \Big(1+ s_{m} \Big)^{2} \bigg]
                                                 \bigg[F_{1}^{2}+ \tau F_{2}^{2} \bigg]\Bigg \} \, , \label{L-SVequal}
\ea 
\ba
 -\textrm{Im}\,(\Pi^{11}+\Pi^{22}) & = &  \frac{ M^{2}}{4 \pi \kappa} \Big( \epsilon^{*}_{F} - \Gamma \Big)
                         \Theta \Big( \epsilon^{*}_{F} - \Gamma \Big)
                      \Bigg \{ 2 \tau \bigg [ F_{1} + \Big( 1+ s_{m} \Big)F_{2} \bigg]^{2} 
+ \Delta \bigg[F_{1}^{2} + \tau F_{2}^{2} \bigg] \Bigg\} \, . \label{T-SVequal}
\ea
We use the notation $s_{m} \equiv S/M$, being $S$ the constant scalar potential, and have introduced the usual 
terms~\cite{Alberico:1988bv,Alberico:1989aja}
\ba
\Gamma \, &=& \, Max  \Bigg\{ \bigg( \epsilon^{*}_{F} - 2 \lambda \Big); \,  
            \Big( -\lambda + \kappa \sqrt{ 1+ \frac{ \big(1+ s_{m}\big)^{2}}{\tau}} \bigg)
                        \Bigg\} \, , \label{Gamma}
                        \\
\Delta \,& = &\,   \frac{ \tau}{\kappa^{2}} \Big( \frac{1}{3} 
                           \big( \epsilon^{*^{2}}_{F} + \epsilon^{*}_{F} \Gamma + \Gamma^{2} \big)
                               + \lambda \big(\epsilon^{*}_{F} + \Gamma \big) +
                              \lambda^{2} \Big) -  \Big( \tau + \big(1+s_{m} \big)^{2} \Big) \, .    \label{Delta}              
\ea
The dimensionless Fermi energy is given by $\epsilon^\ast_F\equiv E^\ast_F/M=\sqrt{(1+s_m)^2+\eta_F^2}$ with $\eta_F\equiv k_F/M$.

For completeness, in this study we also consider the case of the Relativistic Plane Wave Impulse Approximation, that is, no scalar neither vector potentials are considered in the final state that is simply described as a free relativistic Dirac wave function. On the contrary, the initial nucleon states are determined by $S(\np_i)$ and $V(\np_i)$. The general expressions for the components of the tensor and the polarization propagator are given in Appendix B. In next section we present the response functions and discuss in detail the results corresponding to the different approaches considered in the work.

%%%%%%%%%%%%%%%%%%%%%%%%%%%%%%%%%%%%%%%%%%%%%%%%%%%%%%%%%%%%%%%%%%%%%%%%%%%%%%%%% 

%\subsection{Scaling and superscaling}

The study of scaling and superscaling in inclusive quasielastic (QE) electron scattering reactions has been presented in detail in previous works~\cite{Alberico:1988bv,Donnelly:1998xg,Maieron:2001it,Amaro:2004bs,Gonzalez-Jimenez:2014eqa,Megias:2016lke,Megias:2016fjk}. Here we simply summarize the main points and focus on those aspects of more relevance to the discussion that follows. The QE $(e,e')$ cross section at medium-to-high values of the momentum transfer $q$ divided by an appropriate single-nucleon cross section (derived from the analysis of the problem based on the Relativistic Fermi Gas (RFG) model) leads to a general function that is almost independent of $q$ and is the same for all nuclear systems. This is known as scaling of first and second kind, respectively. The occurrence of both types of scaling is denoted as superscaling. From the analysis of the longitudinal response function extracted from $(e,e')$ data, we introduce a general, universal, scaling function that only depends on a single variable $\psi$, known as scaling variable. This is basically given in terms of the minimum nucleon kinetic energy allowed in the process scaled by the Fermi kinetic energy. In the simple case of the RFG model, the scaling function is simply given as $f(\psi)=(3/4)(1-\psi^2)\Theta(1-\psi^2)$. This result clearly differs from the analysis of data that leads to a phenomenological scaling function extended to higher/lower values of $\psi$ with an asymmetric shape and a tail extended to higher positive values of $\psi$. This asymmetry of the scaling function is successfully reproduced by the predictions of the RMF model~\cite{Caballero:2005sj,Caballero:2006wi,Caballero:2007tz}.

In this work our aim is not to get scaling functions from a comparison with data, but to provide a systematic analysis of the scaling and superscaling phenomenon within the framework of the relativistic mean field approach in nuclear matter. Our interest is to analyze the effects introduced by the use of scalar and vector potentials considering both momentum-independent and momentum-dependent potentials. Whereas analytical expressions for the scaling functions can be obtained in the former, for energy-dependent potentials the equations must be solved numerically. Notwithstanding a general discussion of results is shown in next section.

From the general expression of the polarization propagator applicable to the case of equal constant scalar and vector potentials in the initial and final states (\ref{L-SVequal},~\ref{T-SVequal}), the longitudinal and transverse electromagnetic nuclear responses can be written in the form:
\ba
R^L &=&  \frac{3{\cal N}}{4M\kappa\eta_F^3}(\epsilon^\ast_F-\Gamma)\Theta(\epsilon_F^\ast-\Gamma)\frac{\kappa^2}{\tau}\Big\{\left[\Delta+\tau+(1+s_m)^2\right](F_1^2+\tau F_2^2)
- \tau \left(F_1+(1+s_m)F_2\right)^2\Big\} \, ,  \nonumber \\ 
& &  \\
R^T&=& \frac{3{\cal N}}{4M\kappa\eta_F^3}(\epsilon^\ast_F-\Gamma)\Theta(\epsilon_F^\ast-\Gamma)
\Big[2\tau\left( F_1+(1+s_m)F_2 \right)^2+\Delta\left(F_1^2+\tau F_2^2\right) \Big] \, . 
\ea
A detailed analysis of the function $\Gamma$ and its specific behavior at different kinematical situations is presented in \cite{Alberico:1988bv}. In this paper our interest is focused on scaling at high momentum transfers where the value of $\Gamma$ is determined by the second choice in (\ref{Gamma}). The response functions are usually expressed by introducing a general dimensionless scaling variable, $\psi^*$, that is given by
\footnote{The scaling variable can be also extended to the case of different scalar and vector potentials in the initial/final 
states. It is given by
\[ \psi^{*} = \frac{1}{\sqrt{\xi_{F}^\ast}}
            \frac{\lambda^\ast (1+s_{m_i}) - \tau^\ast\rho^\ast}
                { \sqrt{ \tau^\ast (\lambda^\ast\rho^{\ast} + 1 +s_{m_i}) 
                  + \kappa \sqrt{\tau^\ast \big(\tau^\ast\rho^{\ast^2}+(1+ s_{m_i})^{2} 
                        \big) }}} \, . 
\]
}
\be
\psi^{*} =\frac{1}{\sqrt{\xi_{F}^\ast}}
            \frac{\lambda (1+s_{m}) - \tau}
                { \sqrt{ \tau (\lambda + 1 +s_m) 
                  + \kappa \sqrt{\tau \big(\tau+(1+ s_{m})^{2} 
                        \big) }}} 
\ee
with $\xi^\ast_{F} \equiv  (E_{F}^\ast - M^\ast)/M = \epsilon^{*}_{F} - (1 +s_{m})$. 
Notice that the only dependence on the potential enters through the term $s_m=S/M$ that accounts for the effective mass of 
the nucleon (only dependent on the scalar potential $S$).
In the particular case of $S=0$ one recovers the usual expression for the scaling variable as given in previous 
works~\cite{Maieron:2001it,Amaro:2004bs}.

In terms of the scaling variable the longitudinal and transverse electromagnetic response functions result:
\ba
R^L &=& \frac{{\cal N}}{M\kappa\eta_F^3}\xi^\ast_F f(\psi^\ast)\frac{\kappa^2}{\tau}\left\{\left[\Delta+\tau+(1+s_{m})^2\right](F_1^2+\tau F_2^2)
-\tau\left(F_1+(1+s_{m})F_2\right)^2\right\}  \\
R^T &=& \frac{{\cal N}}{M\kappa\eta_F^3}\xi^\ast_F f(\psi^\ast)\left\{2\tau\left[F_1+(1+s_{m})F_2\right]^2+
\Delta(F_1^2+\tau F_2^2)\right\}
\ea
with $\eta_F\equiv k_F/M$ and ${\cal N}$ representing the number of nucleons. The scaling function is given as
\be
f(\psi^\ast)=\frac{3}{4}(1-\psi^{\ast^2})\Theta(1-\psi^{\ast^2}) \, . 
\ee
Although not explicitly written, the previous expressions for the nuclear response functions, that hold separately for protons and neutrons,  incorporate both contributions as shown in (\ref{RL},~\ref{RT}). Furthermore, they 
can be rewritten by introducing the electric and magnetic form factors of the nucleon, i.e., $G_E$ and $G_M$. However, in order to illustrate how the response functions behave with scaling and superscaling, it is more convenient to change notation by referring all kinematical variables to the effective nucleon mass, $M^\ast\equiv M+S=M(1+s_m)$, instead of $M$. Then, we introduce the following dimensionless variables:
\ba
&&
\widetilde{\tau}\equiv |Q^2|/(4M^{\ast^2})=\tau/(1+s_m)^2 \,\, ;\, \, 
\widetilde{\kappa}\equiv q/(2M^\ast)=\kappa/(1+s_m)\,\, ;\,\, 
\widetilde{\eta}_F\equiv k_F/M^\ast=\eta_F/(1+s_m) \nonumber \\
&&
\widetilde{\epsilon}_F\equiv E_F^\ast/M^\ast = \epsilon_F^\ast/(1+s_m)  \,\, ; \,\,\,
\widetilde{\xi}_F \equiv (E_{F}^\ast - M^\ast)/M^\ast  =\widetilde{\epsilon}_F-1 =\xi_F^\ast/(1+s_m)
\nonumber \, .
\ea
From the Dirac and Pauli form factors we can define the "effective" electric and magnetic nucleon form factors as: 
\be
\widetilde{G}_M\equiv F_1+(1+s_m)F_2 \,\,  ; \,\,\,\, \widetilde{G}_E\equiv F_1-\tilde{\tau}(1+s_m)F_2 \, .
\ee
Notice that $F_1$ and $F_2$ are simply given by
\be
F_1=\frac{\widetilde{G}_E+\widetilde{\tau}\widetilde{G}_M}{1+\widetilde{\tau}} \,\, ; \,\,\,\,
(1+s_m)F_2=\frac{\widetilde{G}_M-\widetilde{G}_E}{1+\widetilde{\tau}}  \, .
\ee
After some algebra, the final expressions for the electromagnetic responses in terms of the modified electric and magnetic nucleon form factors result,
\ba
R^L &=& \frac{{\cal N}}{M^{\ast}\widetilde{\kappa}\widetilde{\eta}_F^3}\widetilde{\xi}_F f(\psi^\ast)\frac{\widetilde{\kappa}^2}{\widetilde{\tau}}\left[
\widetilde{G}_E^2+\frac{\widetilde{G}_E^2+\widetilde{\tau}\widetilde{G}_M^2}{1+\widetilde{\tau}}\widetilde{\Delta}\right] 
\label{RL-eff}\\
R^T &=& \frac{{\cal N}}{M^\ast\widetilde{\kappa}\widetilde{\eta}_F^3}\widetilde{\xi}_F f(\psi^\ast) \left[
2\widetilde{\tau}\widetilde{G}_M^2+\frac{\widetilde{G}_E^2+\widetilde{\tau}\widetilde{G}_M^2}{1+\widetilde{\tau}}\widetilde{\Delta} \right] \, ,
\label{RT-eff}
\ea 
where, by analogy with the previous variables, we have introduced $\widetilde{\Delta}\equiv (M/M^\ast)^2\Delta=\Delta/(1+s_m)^2$.
Notice that the two expressions for the response functions are formally identical to the ones obtained in the case of the RFG model~\cite{Alberico:1988bv,Maieron:2001it,Donnelly:1998xg}. This means that in the case of constant scalar and vector potentials being the same in the initial and final states, superscaling is fulfilled with the scaling function being identical to the RFG case but with all variables referred to the effective mass, $M^\ast$, and modifying accordingly the definition of the nucleon form factors. In particular, it is important to point out how the relative contribution of the form factor $F_2$ is modified by the presence of the scalar potential. 

In next section we present a detailed study of our results for the nuclear responses and scaling functions. In the latter one may distinguish between the global scaling function, evaluated dividing the total cross section by the appropriate single-nucleon $eN$ elastic cross section, and the separate longitudinal ($L$) and transverse ($T$) contributions. These are given as the ratio between the corresponding nuclear responses and the appropriate longitudinal (transverse) contributions to the elastic single-nucleon cross section: 
$f_{L,T}\equiv k_F R^{L,T}/G_{L,T}$. Within the RFG model the so-called scaling of zeroth kind, i.e., $f=f_L=f_T$, is exactly fulfilled. 
For a detailed study on this topic and the explicit expressions of $G_{L,T}$ the reader can consult~\cite{Maieron:2001it}.

%%%%%%%%%%%%%%%%%%%%%%%%%%%%%%%%%%%%%%%%%%%%%%%%%%%%%%%%%%%%%%%%%%%%%%%%%%%%%%%%%%%%%%%%
%%%%%%%%%%%%%%%%%%%%%%%%%%%%%%%%%%%%%%%%%%%%%%%%%%%%%%%%%%%%%%%%%%%%%%%%%%%%%%%%%%%%%%%%

\section{Analysis of results}

In this section we present and discuss in detail the results obtained for the nuclear electromagnetic responses and scaling functions corresponding to four nuclear systems: $^{12}$C,  $^{16}$O, $^{40}$Ca and $^{56}$Fe. In all the cases we analyze the different approaches discussed in previous sections, i.e., the relativistic mean field with energy-dependent scalar and vector potentials compared with the case of constant potentials, RPWIA and the pure RFG model taken as reference. We show results for different kinematical situations covering from intermediate momentum transfer values, $q=0.5$ GeV/c, up to 2 GeV/c.

%%%%%%%%%%%%%%%%%%%%%%%%
%%	RESPONSE FUNCTIONS
%%%%%%%%%%%%%%%%%%%%%%%

\begin{figure}
\begin{center}
\includegraphics[scale=0.9]{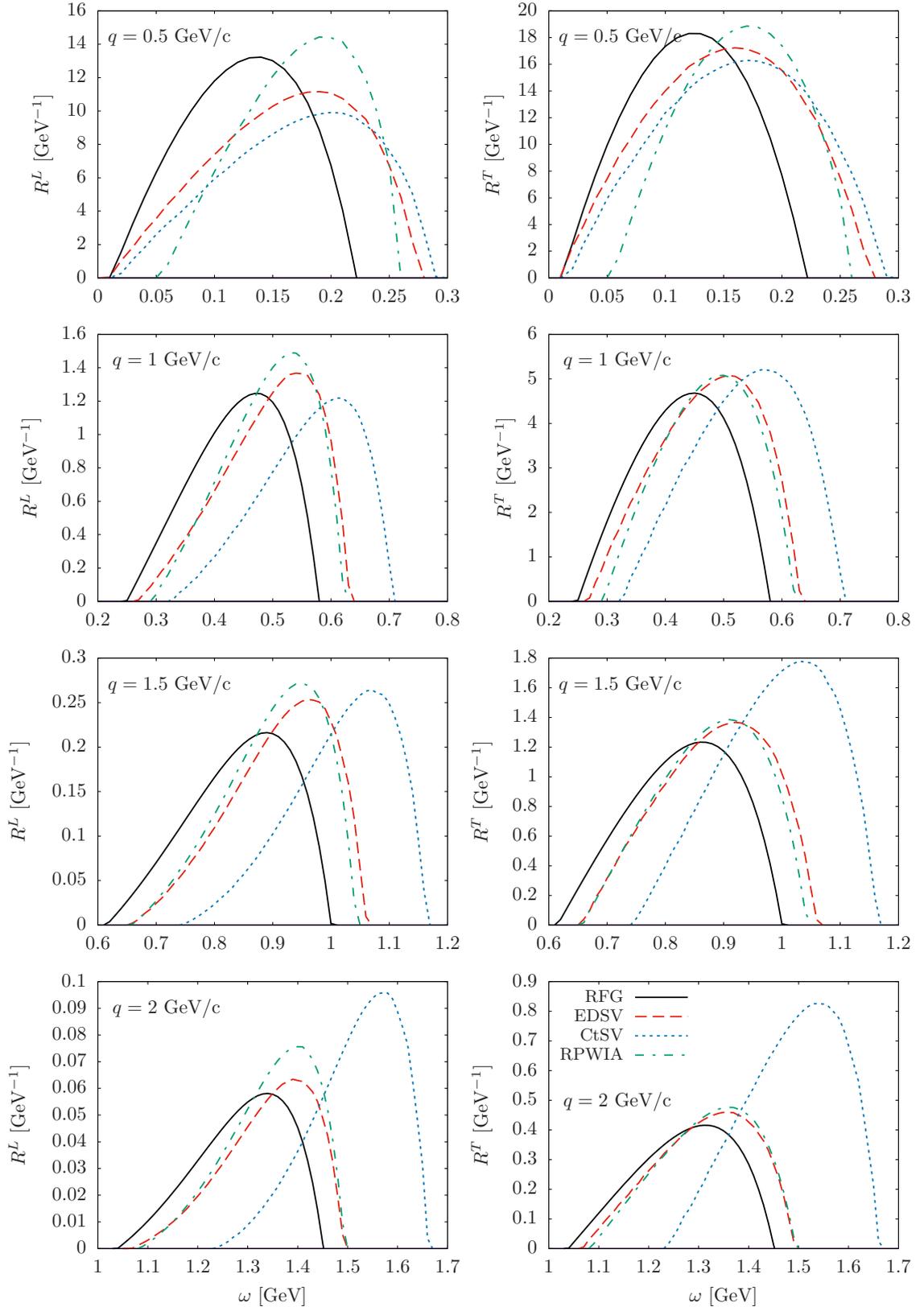}
\caption{Longitudinal (left panels) and transverse (right) 
electromagnetic response functions for $^{12}$C versus the energy transfer $\omega$. Results are shown for different values of the momentum transfer $q$ and the models considered in the work (see text for details): RFG (black solid), EDSV (red dashed), CtSV (blue dotted) and RPWIA (green dot-dashed).} 
\label{Responses_C12}
\end{center}
\end{figure}                                                                      

\begin{figure}
\begin{center}
\includegraphics[scale=0.9]{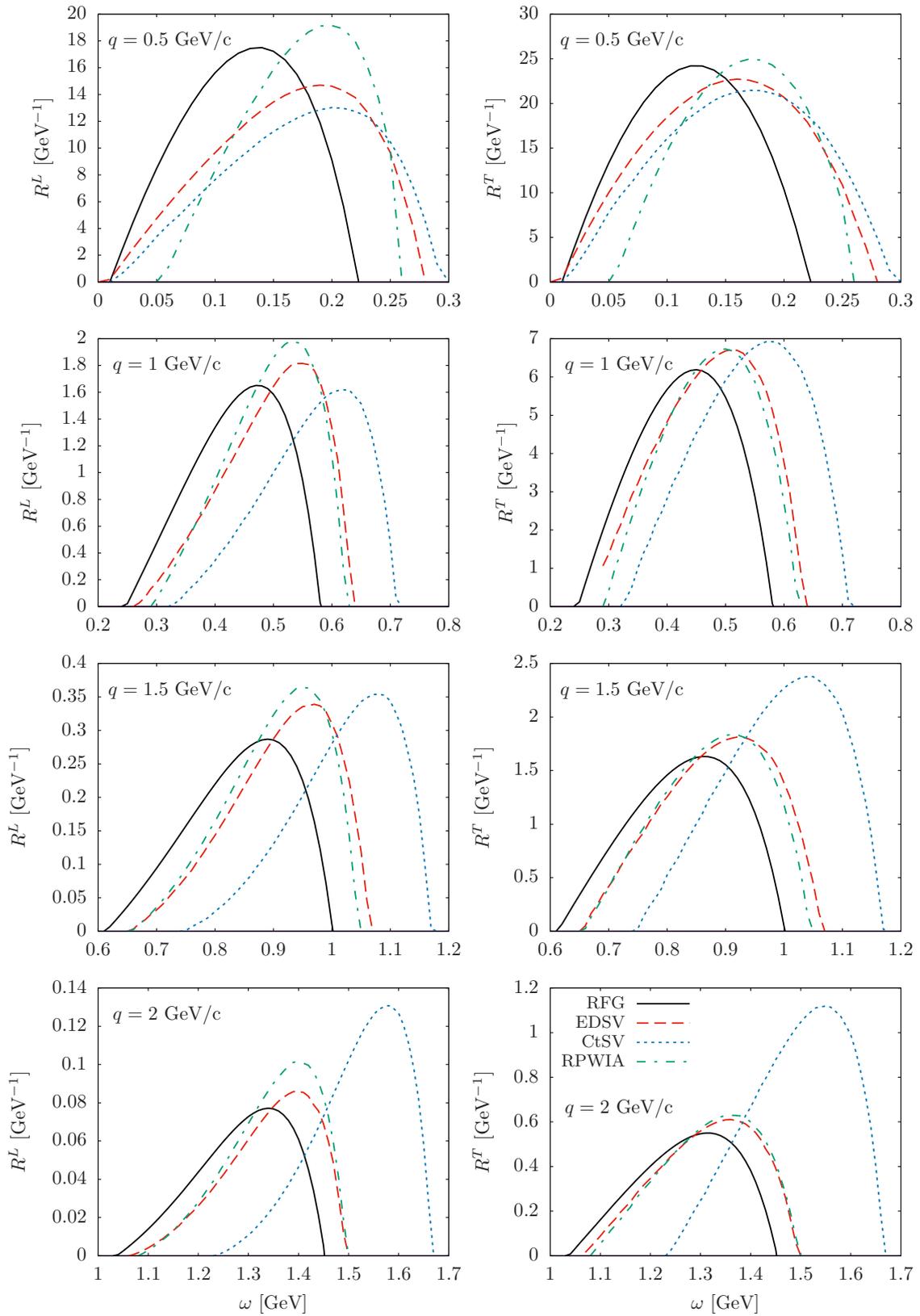}
\caption{Same as Fig.~\ref{Responses_C12}, but for $^{16}$O.}
\label{Responses_O16}
\end{center}
\end{figure}

\begin{figure}
\begin{center}
\includegraphics[scale=0.9]{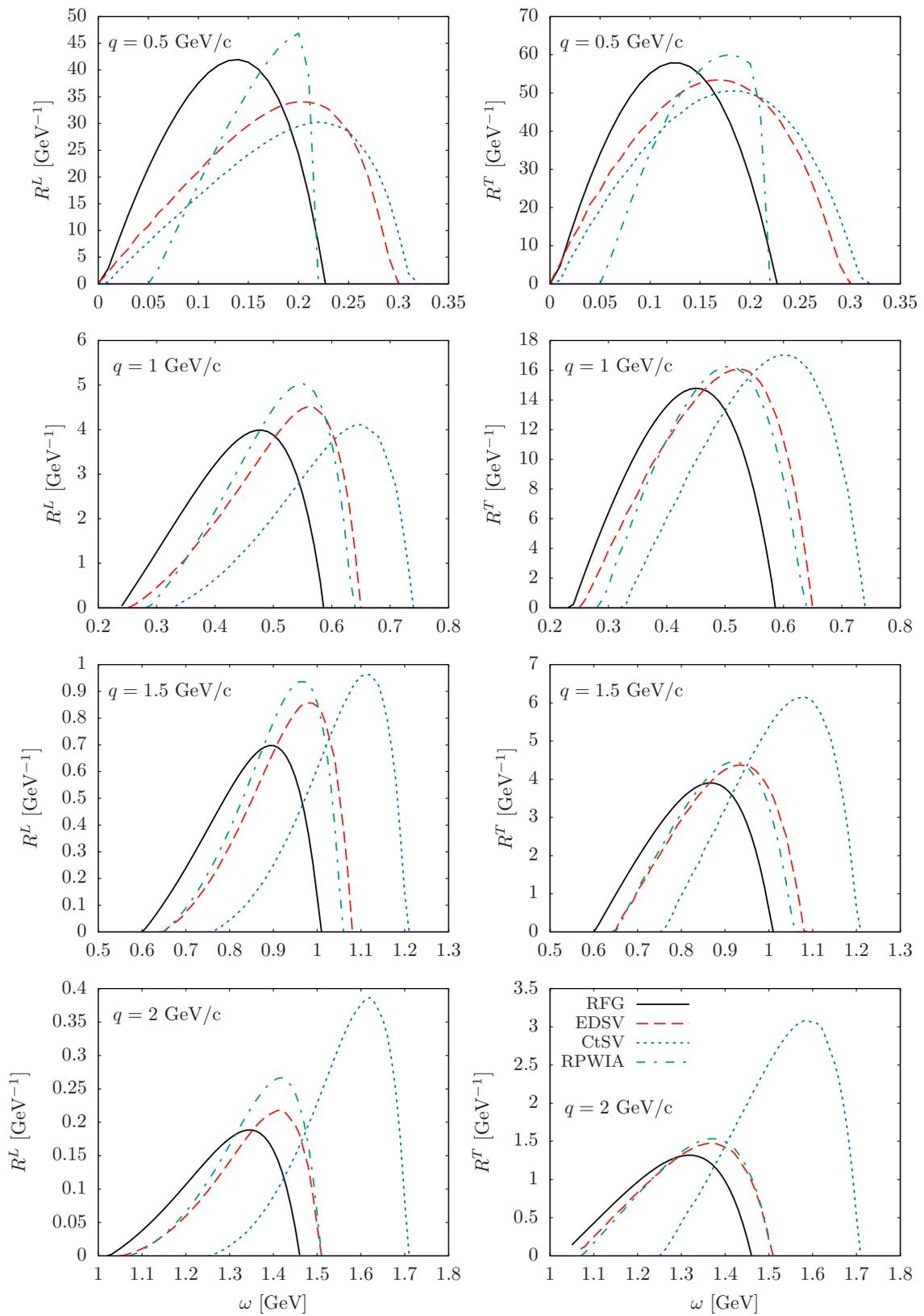}
\caption{Same as Fig.~\ref{Responses_C12}, but for $^{40}$Ca}
\label{Responses_Ca40}
\end{center}
\end{figure}

\begin{figure}
\begin{center}
\includegraphics[scale=0.9]{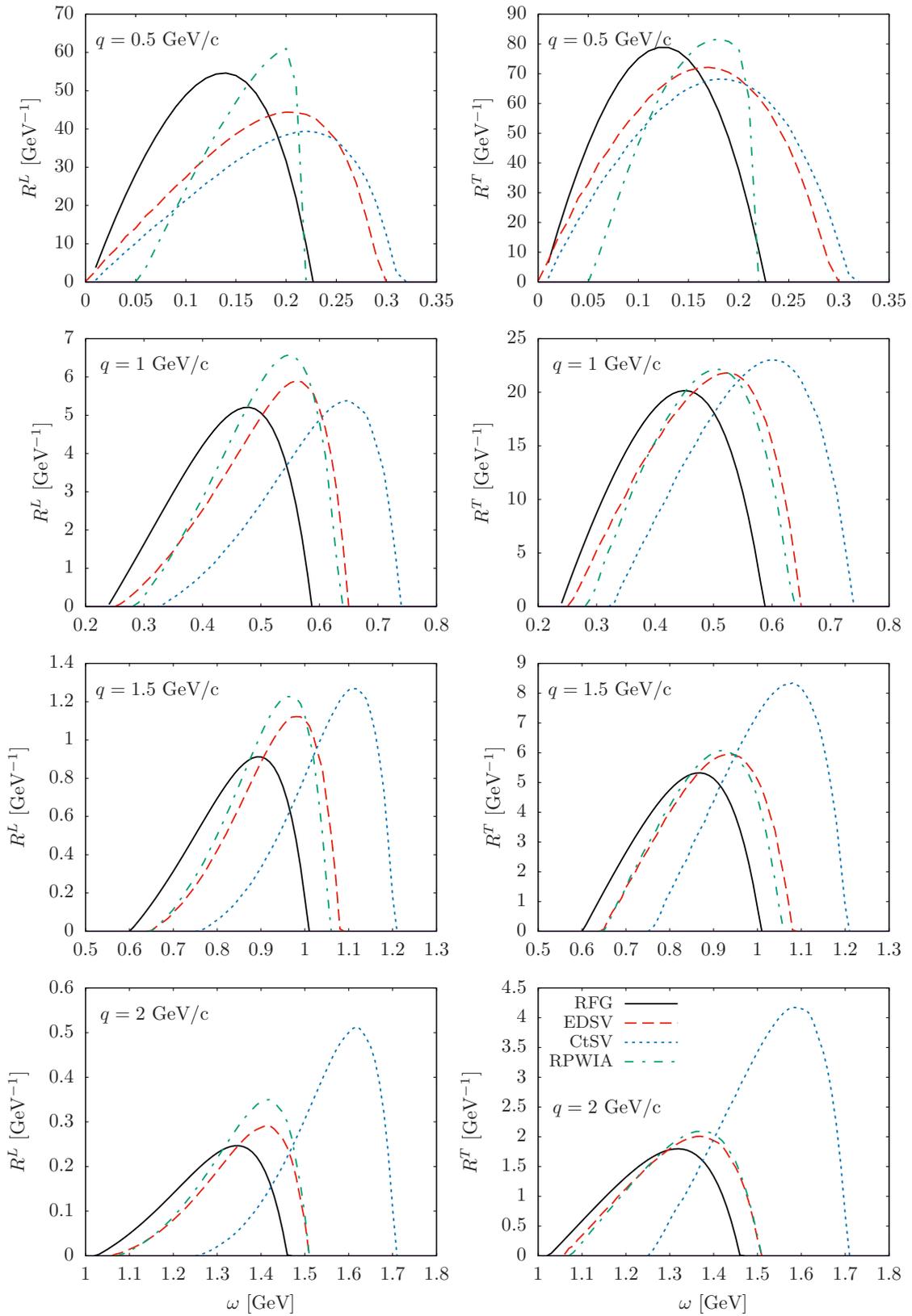}
\caption{Same as Fig.~\ref{Responses_C12}, but for $^{56}$Fe}
\label{Responses_Fe56}
\end{center}
\end{figure}

\begin{figure}
\begin{center}
\includegraphics[scale=0.9]{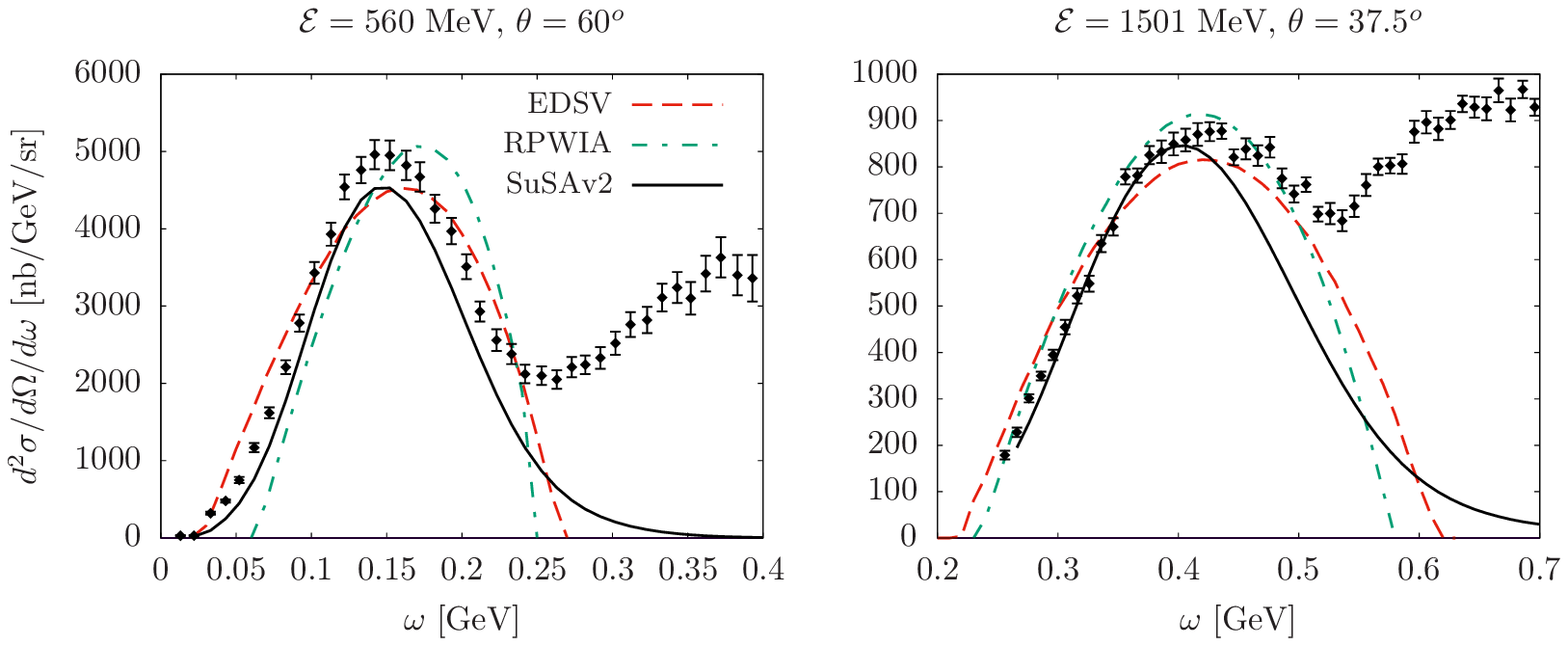}
\caption{Double differential inclusive $^{12}$C$(e,e')$ cross section versus the energy transfer. Predictions corresponding to the EDSV, RPWIA and SuSAv2 approaches are compared with data~\cite{Benhar:2006wy,Benhar:2006er}.}
\label{Cross_section_C12}
\end{center}
\end{figure}

\subsection{Response Functions}

In Fig.~\ref{Responses_C12}
we present the response functions corresponding to $^{12}$C. The value of the Fermi momentum has been fixed to $k_F=228$ MeV/c.
Panels on the left (right) refer to the longitudinal (transverse) responses that are shown in terms of the transfer energy, $\omega$, for fixed values of $q$: 0.5 GeV/c (top panels), 1 GeV/c (second row), 1.5 GeV/c (third) and 2 GeV/c (bottom). In each case we compare the predictions given by the pure RFG (black solid line), RPWIA (green dot-dashed), constant scalar and vector potentials in both the initial and final nucleon states, denoted as CtSV (blue dotted) and energy-dependent scalar and vector potentials, named EDSV (red dashed). 

As observed, the role introduced by the potentials gives rise to very significant effects that depend on the particular approach considered and the response, longitudinal/transverse, analyzed. At $q=0.5$ GeV/c (top panels), the results corresponding to scalar and vector potentials taken as constant or including energy dependence are not so different. Compared to the pure RFG, both responses are extended to larger $\omega$-values, getting smaller values for the maximum and showing more asymmetry. This is particularly true in the case of the longitudinal response. The plane wave limit approach (RPWIA) differs clearly from the other models. 
Notice that the minimum value of $\omega$ allowed by the kinematics is significantly larger. This is connected with the role played by the potentials only in the initial state (see previous sections for details). On the contrary, the maximum $\omega$-value allowed is located between the pure RFG result and the ones corresponding to the models where the potentials enter in both the initial and final states: CtSV and EDSV. The shape of the RPWIA response also differs significantly from the other approaches. 

For higher $q$-values, the results that depart the most are the ones corresponding to the CtSV approach. Notice that this discrepancy gets larger as $q$ increases. On the contrary, the results for the RPWIA and EDSV models get closer, particularly in the case of the transverse response. This is consistent with the kinematics selected where much larger values of the transfer energy and momentum (likewise for the ejected nucleon momentum/energy) are involved. This means that the strength of the potentials in the final state are much weaker. Contrary to the discussion for $q=0.5$ GeV/c, at higher $q$ the role of the potentials gives rise to responses with maxima larger than the pure RFG model. 

In Figs.~\ref{Responses_O16}, \ref{Responses_Ca40}, \ref{Responses_Fe56} we present the electromagnetic responses obtained with the different approaches for $^{16}$O, $^{40}$Ca and $^{56}$Fe, respectively. The values used for the Fermi momentum are $k_F=230$ MeV/c ($^{16}$O) and 241 MeV/c ($^{40}$Ca and $^{56}$Fe).  As observed, the results show a similar behavior to the ones for $^{12}$C except for the magnitude of the responses, and the discrepancy introduced by the different models follows a similar trend to the one discussed in Fig.~\ref{Responses_C12}. In all cases it is remarkable the accordance between the RPWIA and EDSV predictions for the transverse responses at $q\geq 1$ GeV/c. For the longitudinal response the difference is clearly visible, RPWIA being higher than the EDSV result. The main discrepancy at high $q$ is introduced by the CtSV model due to the strong scalar and vector constant potentials involved in both the initial and final states. 
Finally, notice the behavior shown by the RPWIA predictions corresponding to the heavier nuclear systems, $^{40}$Ca and $^{56}$Fe, at $q=0.5$ GeV/c. This is simply a consequence of the assumptions implied by the RPWIA model that lead the value of $\tau^\ast$ to become negative at lower $q$ (0.5 GeV/c) and larger $k_F$-values ($^{40}$Ca and $^{56}$Fe).

To conclude we present in Fig.~\ref{Cross_section_C12} the double differential inclusive $(e,e')$ cross sections for $^{12}$C against the energy transfer $\omega$. Two kinematical situations for fixed values of the electron beam energy (${\cal E}$) and scattering angle ($\theta$) have been selected. The panel on the left, ${\cal E}=560$ MeV, $\theta=60^o$, corresponds to a value of the momentum transfer at the quasielastic peak, namely, the value of $q$ where the maximum in the QE peak appears, $q_{QE}=508$ MeV/c. On the contrary, the panel on the right (${\cal E}=1501$ MeV, $\theta=37.5^o$) corresponds to $q_{QE}=917$ MeV/c. These values are consistent with the ones considered in the previous analysis of the response functions. In each panel we confront the predictions of the EDSV and RPWIA models with the elaborate SuSAv2 calculation~\cite{Megias:2016lke} and with the available experimental data~\cite{Benhar:2006wy,Benhar:2006er}. As observed, the EDSV prediction for the maximum agrees better with SuSAv2 and data, while the RPWIA results are too high. Notwithstanding significant discrepancies are clearly visible between the model predictions due to the approaches implicit in nuclear matter calculations. Whereas SuSAv2, based on RMF applied to finite nuclei, extends to low and high $\omega$ values, the EDSV and RPWIA calculations, based on nuclear matter, show strict limits in $\omega$ due to the fixed Fermi momentum. Notice also that EDSV and RPWIA cross sections are significantly larger than the SuSAv2 prediction for $\omega$-values on the right of the maximum of the QE peak. However, the EDSV model, in spite of its simplicity, is in accordance with data in the QE domain.

%%%%%%%%%%%%%%%%%%%%%%%%%%%%%%%%%%%%%%%%%%%%%%%%%%%%%%%%%%%%%%%%%%%%%%%%
%%	SCALING FUNCTIONS
%%%%%%%%%%%%%%%%%%%%%%%%%%%%%%%%%%%%

\begin{figure*}
\begin{center}
\includegraphics{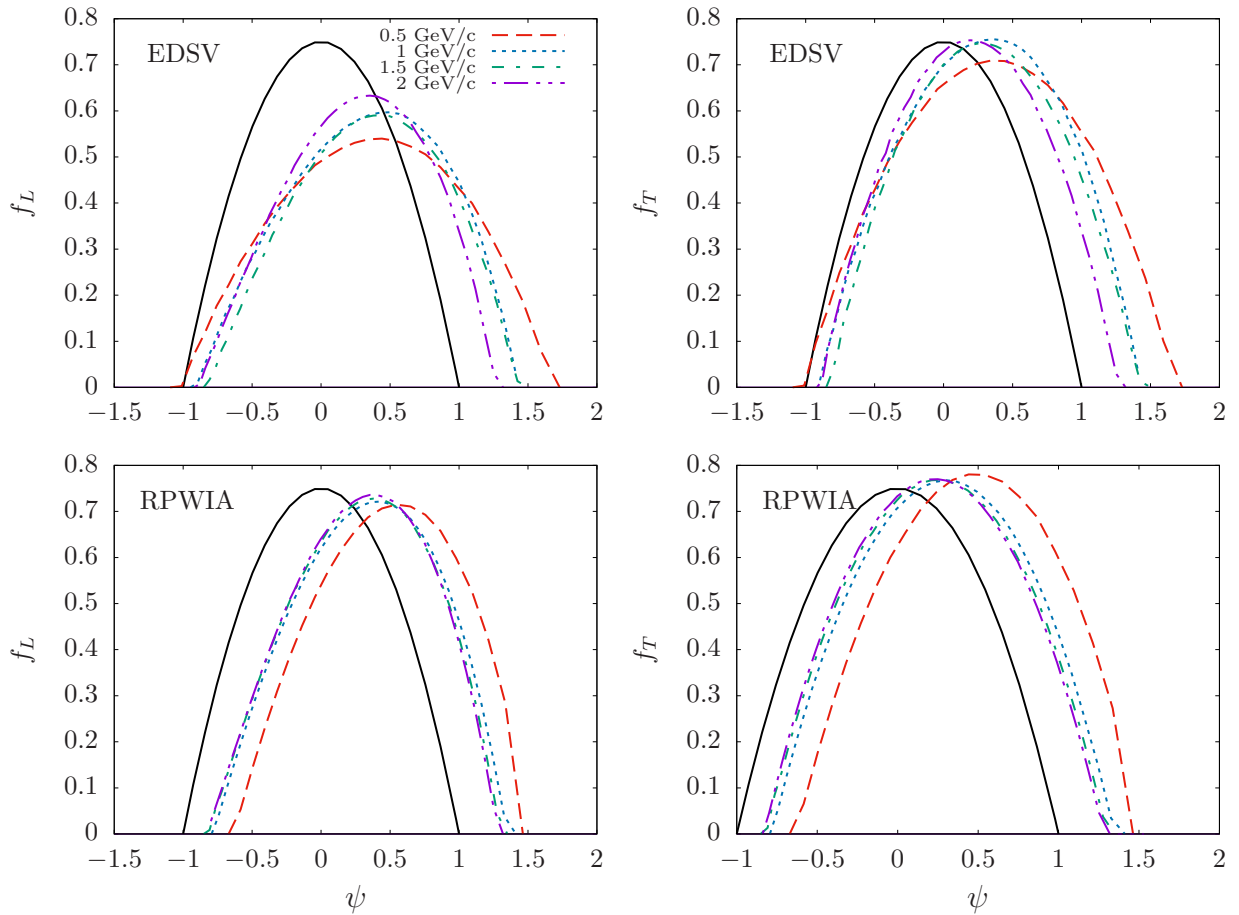}
\caption{Longitudinal (left panels) and transverse (right) scaling functions for $^{12}$C. Top (bottom) panels correspond to the EDSV (RPWIA) models. Results are presented for different values of the momentum transfer: $q=0.5$ GeV/c (red dashed), $q=1$ GeV/c (blue dotted), $q=1.5$ GeV/c (green dot-dashed) and $q=2$ GeV/c (magenta double-dotted dashed). The RFG prediction (black solid) is also shown for reference.} 
\label{scaling-first}
\end{center}
\end{figure*}

\begin{figure*}
\begin{center}
\includegraphics{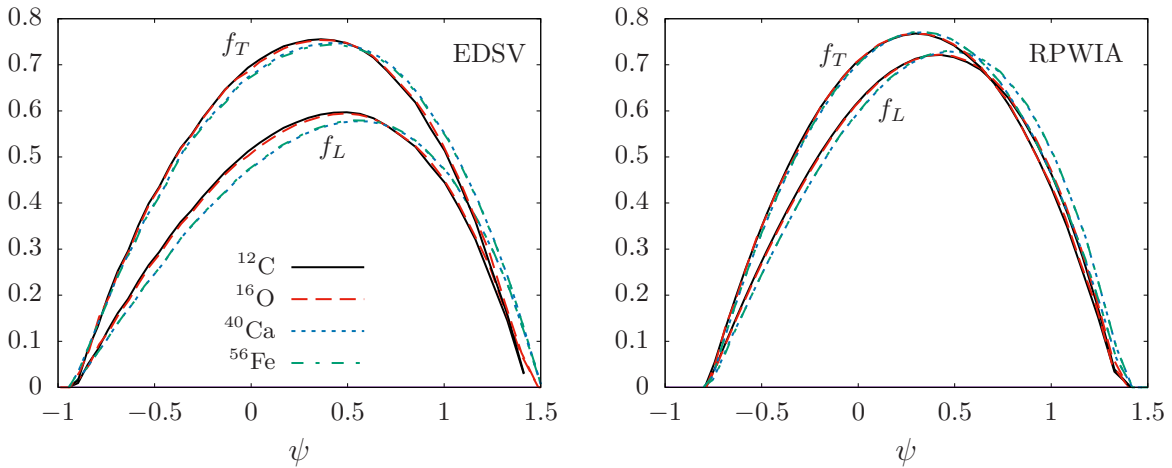}
\caption{Longitudinal and transverse scaling functions evaluated with the EDSV model (left panel) and RPWIA (right) for the four nuclear systems considered: $^{12}$C (black solid), $^{16}$O (red dashed), $^{40}$Ca (blue dotted) and $^{56}$Fe (green dot-dashed). In the two panels, $f_L$ ($f_T$) corresponds to the lower (upper) curves.} 
\label{scaling-second}
\end{center}
\end{figure*}

\begin{figure*}
\begin{center}
\includegraphics{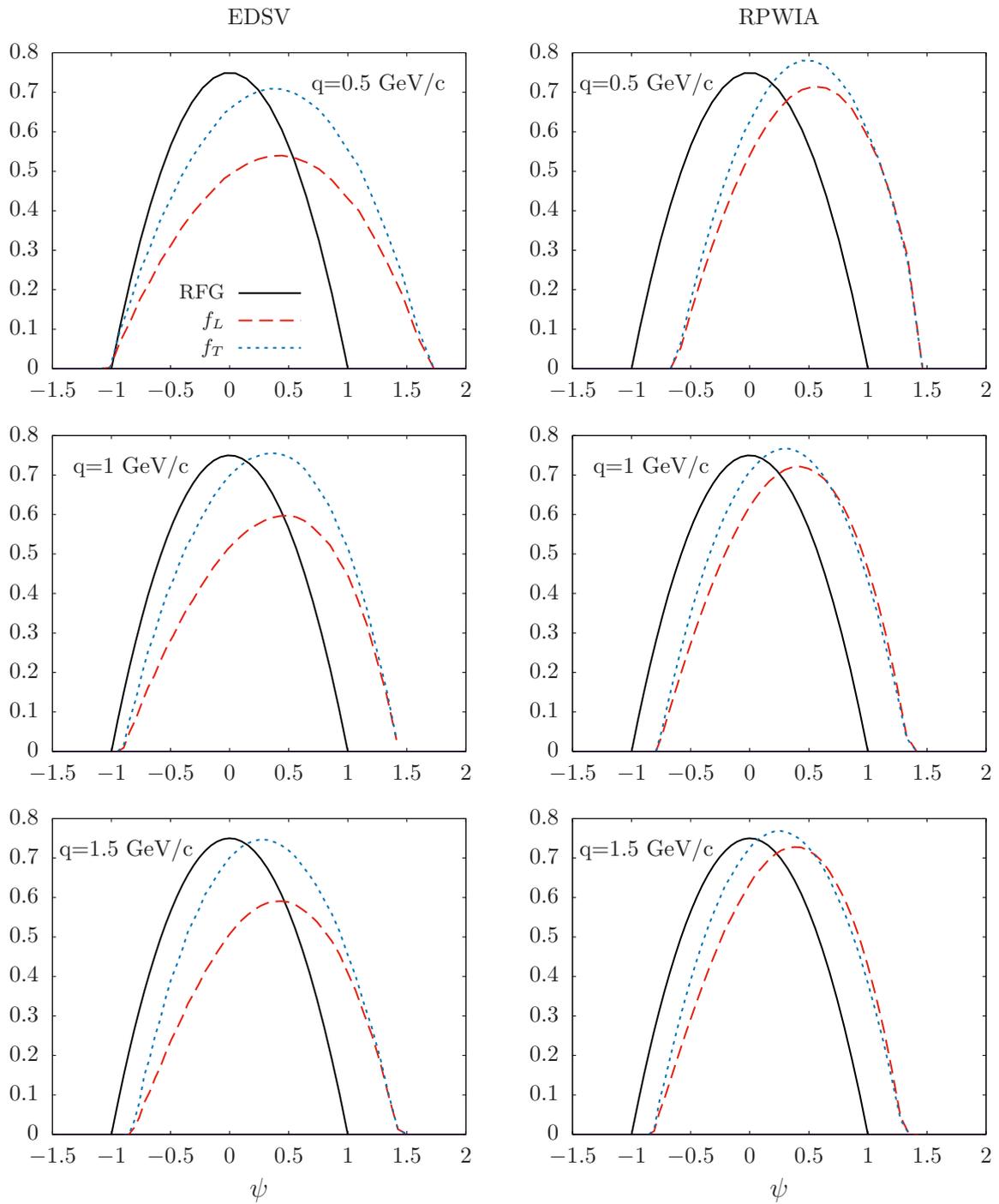}
\caption{Scaling functions for $^{12}$C corresponding to different $q$-values. Left (right) panels correspond to the EDSV (RPWIA) models and the separate longitudinal (red dashed) and transverse (blue dotted) functions are shown. The RFG prediction (black solid) is presented for reference.} 
\label{scaling-zeroth}
\end{center}
\end{figure*}

%\begin{figure}
%\begin{center}
%\includegraphics[h]{fLfTctegra.eps}
%\caption{Analisis of Zero-kind scaling.  Global scaling function $fL$ 
%and $fT$ compare with separate $L$, $T$, contribution.  RMF compared with 
%$S$, $V$, $cte$ model (Left panel), and $RPWIA$ with $S$, $V$, $cte$ 
%model (right panel). Nucleus 12C}
%\label{scaling-zeroth-2}
%\end{center}
%\end{figure}

\begin{figure*}
\begin{center}
\includegraphics{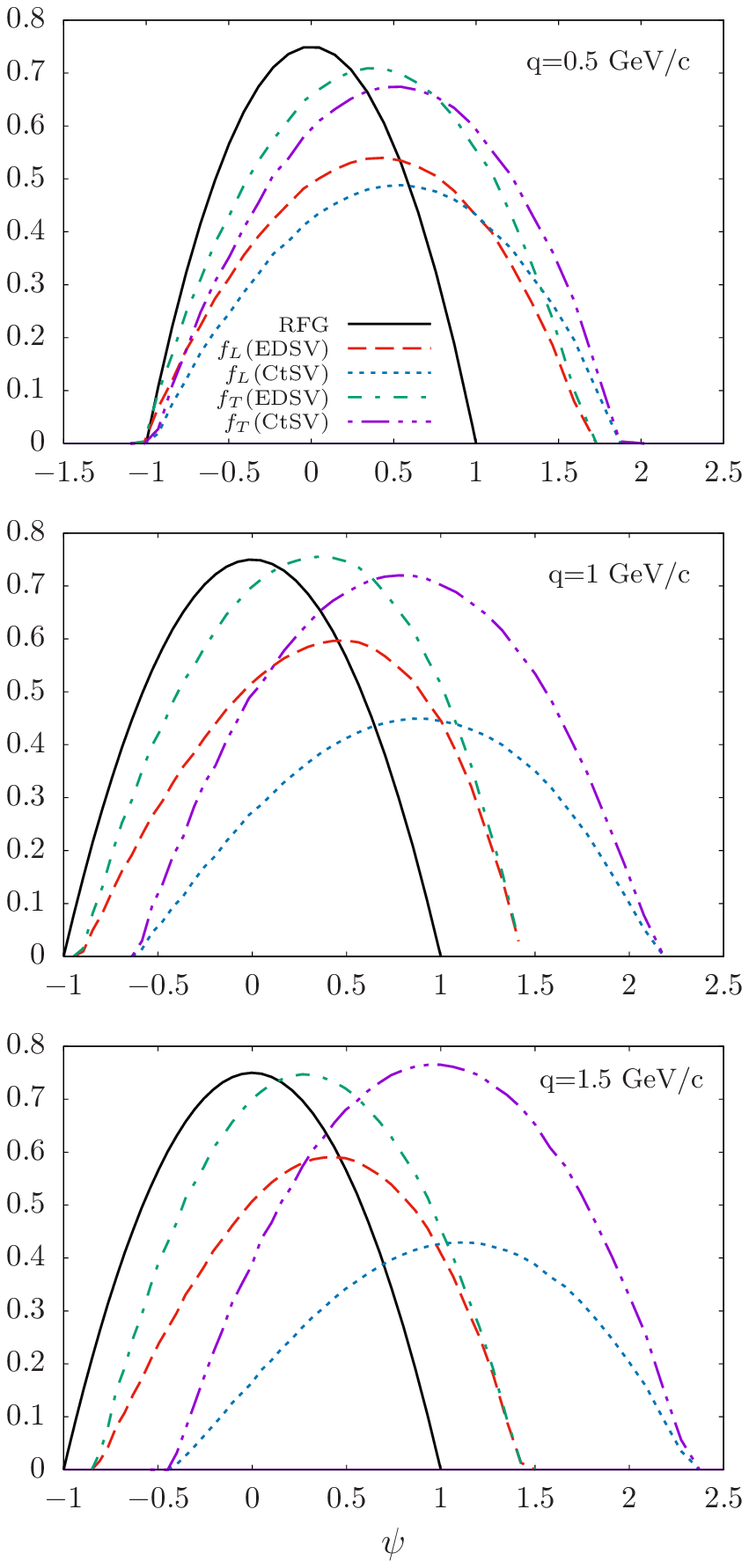}
\caption{Longitudinal and transverse scaling functions for $^{12}$C at different $q$-values. The different curves correspond to: i) $f_L$ evaluated for the EDSV model (red dashed) and CtSV (blue dotted), and ii) $f_T$ for EDSV (green dot-dashed) and CtSV (magenta double-dotted dashed). The RFG prediction (black solid) is also shown for reference.}
\label{scaling-zeroth-3}
\end{center}
\end{figure*}

%%%%%%%%%%%%%%%%%%%%%%%%%%%%%%%%%%%%%%%%%%%%%%%%%%%%%%%%%%%%%%%%%%%%%%%%%
\subsection{Scaling Functions}

The analysis of scaling and superscaling is shown in Figs.~\ref{scaling-first}-\ref{scaling-zeroth-3}. In what follows we discuss the results corresponding to different kinematics and several nuclei. We present the results for the superscaling function as given in previous sections for the several approaches considered in this work. In fig.~\ref{scaling-first} we analyze scaling of first kind, i.e., the scaling function's independence of the momentum transfer $q$. We have selected the EDSV model and the plane wave limit (RPWIA), and show results only for $^{12}$C. The superscaling function corresponding to the RFG , i.e., $(3/4)(1-\psi^2)\Theta(1-\psi^2)$, is also shown for reference. Top panels contain the results corresponding to the case of energy-dependent scalar and vector potentials in the initial and final states, whereas bottom panels refer to the RPWIA approach. We show separately the longitudinal (panels on the left) and transverse (right) scaling functions comparing the predictions corresponding to four different values of the momentum transfer, $q=0.5$ GeV/c (red dashed), 1 GeV/c (blue dotted), 1.5 GeV/c (green dot-dashed) and 2 GeV/c (magenta double-dotted dashed). 

As observed, a clear difference occurs between the two models. In the RPWIA case (bottom panels) the shape and magnitude of the scaling functions are similar to the RFG ones, except for a shift in the scaling variable (shift in the transfer energy $\omega$). Furthermore, scaling of first kind works extremely well for $q\geq 1$ GeV/c. Only the case $q=0.5$ GeV/c departs from the others being shifted to larger $\psi$-values. Finally, notice that both longitudinal and transverse scaling functions are similar. On the contrary, the results obtained within the EDSV model for $f_L$ and $f_T$ clearly differ being for the latter much higher (with the peaks close to 0.75, the maximum value reached by the RFG result). In the case of $f_L$ the maxima are below $\sim 0.6$ (about $20\%$ lower than the RFG peak). Furthermore, contrary to RPWIA, scaling of first kind breaks down with the maxima decreasing and the tails being more extended to larger $\psi$-values as the momentum transfer gets smaller. The magnitude of this effect depends on the particular $q$-values selected. Notice that the scaling functions for $q=1$ and 1.5 GeV/c are in very good accordance. It is important to point out that the significant reduction observed in the longitudinal scaling function reflects the variation introduced in the nucleon form factors due to the presence of the scalar potential and consequently the modification ascribed to the effective mass.
%{\ttred Not very sure of the previous sentence in blue}.

In Fig.~\ref{scaling-second} we analyze scaling of second kind by showing the scaling functions obtained for carbon, oxygen, calcium and iron. As in the previous case, we consider the EDSV model (left panel) and RPWIA (right panel). Each panel contains the separate longitudinal (lower curves) and transverse (upper curves) scaling functions. All results correspond to $q=1$ GeV/c.
Notice that scaling of second kind, i.e., independence of the nuclei, works perfectly well. All results collapse into a single (very narrow) scaling function. This is consistent with the analysis of data. However, note that the magnitude of the functions is very different in each case, in particular, for $f_L$ as compared to $f_T$ when evaluated with the EDSV model. 
For completeness in Fig.~\ref{scaling-zeroth} we compare directly the longitudinal and transverse scaling functions for the two models considered and three different values of the momentum transfer. The RFG prediction is also shown for reference. In the RPWIA case (right panels), as already mentioned, the results are not very different to the RFG ones, apart from the shift in the scaling variable. The difference (in the maximum) between $f_L$ (red dashed) and $f_T$  (blue dotted) is of the order of $\sim 8-10\%$ with $f_L$ being smaller. On the contrary, the discrepancies are much more significant in the case of the EDSV approach where $f_L$ departs from $f_T$ by $\sim 30-35\%$ (at $q=0.5$ GeV/c) and $\sim 20-25\%$ ($q=1$ and 1.5 GeV/c). Whereas $f_T$ reaches its maximum close to the RFG result (0.75), $f_L$ is strongly reduced. 

To conclude, in Fig.~\ref{scaling-zeroth-3} we compare the EDSV predictions for $f_L$ (red dashed) and $f_T$ (green dot-dashed) with the results evaluated using constant scalar and vector potentials in the initial and final states (CtSV): $f_L$ (blue dotted) and $f_T$ (magenta double-dotted dashed). As in the previous figure we consider three kinematical situations corresponding to $q=0.5$, 1 and 1.5 GeV/c. The RFG prediction is incorporated for reference. Only at $q=0.5$ GeV/c the predictions of CtSV and EDSV are similar for both the longitudinal and transverse scaling functions, although the former much smaller (the maxima reached at $\sim 0.5$). For higher $q$-values the discrepancy between $f_L$ and $f_T$ remains for both models, although the CtSV predictions are extended to much larger $\psi$-values. This result is already observed within the RMF model applied to finite nuclei, and it is connected with the strength of the scalar and vector potentials in the final state.
%It is important to point out that all these scaling functions, longitudinal and transverse, corresponding to the case of constant scalar and vector potentials, collapse into a unique (universal) scaling function (the same as in the RFG model) when represented versus the modified scaling variable taking into account the effective mass introduced by the scalar potential (\ref{RL-eff},~\ref{RT-eff}).

%{\ttred What else???}

%%%%%%%%%%%%%%%%%%%%%%%%%%%%%%%%%%%%%%%%%%%%%%%%%%%%%%%%%%%%%%%%%%%%%%%%%%%%%%%%
%%%%%%%%%%%%%%%%%%%%%%%%%%%%%%%%%%%%%%%%%%%%%%%%%%%%%%%%%%%%%%%%%%%%%%%%%%%%%%%%%

\section{CONCLUSIONS}

In this paper we have presented a systematic study of the electromagnetic quasielastic $(e,e')$ response functions within a relativistic mean field applied to nuclear matter. Our goal has been to explore the effects introduced by the scalar and vector potentials in both the initial and final nucleon states. Although being aware of the over-simplified description provided by the model, it allows us to get analytical expressions for the responses whose behavior can be then explored in detail. Emphasis has been placed on the effects linked to the particular structure of the scalar and vector potentials used. Different options have been considered; from energy/momentum dependent potentials to fixed-constant ones and the special case of the plane wave limit, i.e., no potential in the final state. The use of relativistic potentials being equal and/or different in the initial and final states has also been analyzed at depth.

As already investigated in the past, the RMF applied to finite nuclei has shown its capability to explain the detailed structure of the $(e,e')$ responses. However, this model clearly fails at very high values of the momentum transfer because of the strong potentials used. This behavior has been clearly proved in the present work, also showing how the responses evaluated with energy-dependent potentials get closer to the plane wave limit predictions as the energy/momentum values increase. This is fully consistent with the assumptions considered in the SuSAv2 model~\cite{Gonzalez-Jimenez:2014eqa,Megias:2016lke,Megias:2017cuh}.

Finally, we have completed our systematic study by incorporating scaling and superscaling arguments. We have computed the separate longitudinal and transverse scaling functions for the different models and several nuclei, and have analyzed how scaling of first (independence on the momentum transferred) and second kind (nucleus independence) work. We have shown the specific role of the relativistic potentials in scaling breakdown. Whereas the scalar term modifies the nucleon mass (an effective mass is introduced), the vector term affects the energy but disappears in the particular case of using identical potentials in both the initial and final states. It is also important to point out that superscaling is perfectly fulfilled under some conditions, i.e., the use of constant scalar/vector potentials being the same in the initial and final states, but with new scaling variables referred to the effective mass and the nucleon form factors modified accordingly.

In summary, our objective in this work has been to explore, within a simple but fully relativistic approach, the effects associated to the use of scalar and vector potentials in the electromagnetic $(e,e')$ responses and in the scaling and superscaling functions. This has allowed us to understand in a rather transparent way the limitations of other more realistic models, i.e., the RMF applied to finite nuclei, when compared with data at very different kinematics.

\section*{Acknowledgements}

This work has been partially supported by the Spanish Ministerio de Ciencia e Innovaci\'on and ERDF (European Regional Development Fund) under Contract No. FIS2017-88410-P, by the Junta de Andalucia (FQM 160, SOMM17/6105/UGR), and by the Spanish Consolider-Ingenio 2000 program CPAN (CSD2007-00042).

\section*{Appendix A.- Functional momentum dependence of the scalar and vector potentials (EDSV model)}

The explicit expressions for the  phenomenological scalar and vector potentials, $S(\np)$ and $V(\np)$, used in this work are as follows: 
\ba
&&S({\bf{p}}) \, = \, \alpha S_{0} \left[a_{0} \, + \, a_{1} T({\bf{p}}) \, + \,  a_{2}T({\bf{p}})^{2}\right]  \nonumber \\
&& V({\bf{p}}) \, = \, \alpha V_{0}\left[ b_{0} \, + \, b_{1} T({\bf{p}}) \, + \, b_{2}T({\bf{p}})^{2}\right]  \nonumber \, ,   
\ea
where, following~\cite{Kim:1994nr,Horowitz:1993nb}, we have used $S_{0} \, = \, -0.431$  GeV and 
$ V_{0} \, = \, 0.354$  GeV for the constant scalar and vector potentials. The term $T({\bf{p}})$ is the  kinetic energy of the nucleon and the  parameters $a_i, b_i$ are given by:
$a_{0} = 0.97$, $a_{1} = -0.66 $, $a_{2} = 0.28$,  $b_{0} = 0.97$, $ b_{1} = -0.97$
and $b_{2} = 0.33$. The factor  $\alpha = k_{F}/k^{0}_{F}$ represents an average over the
nuclear volume with $k^{0}_{F} = 0.257$ GeV/c the standard value of the Fermi momentum for nuclear matter.

\section*{Appendix B.- Relativistic Plane Wave Impulse Approximation (RPWIA)}
 
In this appendix we present the explicit expressions obtained in the Relativistic Plane Wave Impulse Approximation. Notice that the scalar and vector potentials only enter in the initial state, i.e., $S(\np_i)$ and $V(\np_i)$. Making use of the general set of dimensionless variables introduced in (\ref{adimension}), the longitudinal and transverse components of the tensor $T^{\mu\nu}$ can be written as,
\ba
T^{00}_{PW} & =   & 2M^{2} \Bigg \{ - \bigg[ \tau^{*} + \frac{s^{2}_{m_{i}}}{4} \bigg]
                               \bigg[F_{1} + \Big( 1+\frac{s_{m_{i}}}{2}\Big)F_{2}\bigg]^{2}
             + \bigg[\epsilon^{*^{2}}_{i} + 2 \lambda^{*} \epsilon^{*}_{i} \bigg]
                                     \bigg[F_{1}^{2} + \tau^{*} F^{2}_{2} \bigg]
                                        \nonumber \\
       &     &   +     2 \lambda^{*}\epsilon^{*}_{i} \bigg[ \frac{-s_{m_{i}}}{2} F_{1}F_{2} +
                                           \Delta m^{*^{2}} F_{2}^{2} \bigg]
	+ \lambda^{*^{2}} \bigg [ \Big(\tau^{*} + 2 \Delta m^{*^{2}} - 
                             \big(1+ \frac{s_{m_{i}}}{2} \big)^2 \Big) F_{2}^{2} -
                                     2\Big( 1 + s_{m_{i}} \Big)F_{1} F_{2} \bigg] \Bigg \} \, ,
                        \nonumber \\
\ea   
\ba
&&T^{11}_{PW}+T^{22}_{PW} =  4 M^{2} \Bigg \{ \bigg[ \tau^{*} + \frac{s^2_{m_{i}}}{4} \bigg]
                                \bigg[F_{1} + \Big( 1+\frac{s_{m_{i}}}{2}\Big) F_{2} \bigg]^{2}
         	 +   \Delta v \bigg[ -\epsilon^{*}_{i} s_{m_{i}} -2 \lambda^{*} 
                                                 \Big(1+s_{m_{i}} \Big) \bigg]F_{1}F_{2}
                                             \nonumber \\
        &     & + \Delta v \bigg[ \Delta v \Big(\epsilon^{*^{2}}_{i} + 
                  2 \epsilon^{*}_{i} \lambda^{*} - \tau^{*} \Big) + 
                  2 \Delta m^{*^{2}} \Big(\epsilon^{*}_{i}+ \lambda^{*} \Big ) -
                  \Big(\Delta v + 2 \lambda^{*} \Big) 
                           \Big(1+ \frac{s_{m_{i}}}{2} \Big)^{2} \bigg]F_{2}^{2}
                                    \nonumber \\
        &    & + \frac{1}{2} \bigg[ \frac{ \tau^{*}}{\kappa^{2}} \Big( \epsilon^{*^{2}}_{i} +
                                2\epsilon^{*}_{i} \lambda^{*} \rho^{*} + \lambda^{*^{2}}
                                  - \Delta m^{*^{2}} \big(1+ \rho^{*} \big) \Big) -
                                       \Big(\tau^{*} + \big(1+ s_{m_{i}} \big)^2 \Big) \bigg]
                                           \bigg[ F^{2}_{1} + \tau F_{2}^{2} \bigg] \Bigg \} \, .
%                         \nonumber \\
\ea
In the particular case of the scalar and vector potentials assumed to be constant the integrals involved in the polarization propagator can be solved analytically and the final results can be written in the following form,
\ba
-\textrm{Im}\,\Pi^{00}_{PW} & = & \frac{M^{2}}{4 \pi \kappa} \bigg( \epsilon^{*}_{F} - \Gamma^{*} \bigg) 
                                 \Theta \bigg(\epsilon^{*}_{F}- \Gamma^{*} \bigg) 
                      \Bigg \{-\bigg[ \tau^{*} + \frac{s^{2}_{m_{i}}}{4} \bigg] 
                                \bigg[F_{1} + \Big(1 +\frac{s_{m_{i}}}{2} \Big)F_{2} \bigg]^{2}
                                       \nonumber \\
              &   &   +\lambda^{*^{2}}\bigg[ 
                      \Big(\tau^{*} + 2 \Delta m^{*^{2}}- 
                       (1+ \frac{s_{m_{i}}}{2})^2\Big)F^{2}_{2}-
                             2\Big(1+s_{m_{i}} \Big)F_{1}F_{2} \bigg] 
                                         \nonumber \\
              &  &   + \lambda^{*} \bigg[ \epsilon^{*}_{F} + \Gamma^{*}\bigg]
                          \bigg[-\frac{s_{m_{i}}}{2}F_{1}F_{2} + \Delta m^{*^{2}}F^{2}_{2} \bigg]
                                  \nonumber \\
             &  & + \bigg[ \frac{1}{3} \Big( \epsilon^{*^{2}}_{F}+ \epsilon^{*}_{F} \Gamma^{*} +
                                                \Gamma^{*^{2}} \Big) +
                                  \lambda^{*} \Big(\epsilon^{*}_{F} + \Gamma^{*} \Big) \bigg]
                              \bigg[F^{2}_{1} + \tau^{*} F^{2}_{2} \bigg] \Bigg\} \, ,
\ea
\ba
-\textrm{Im}\,\left(\Pi^{11}_{PW}+\Pi^{22}_{PW}\right) & = & \frac{M^{2}}{4 \pi \kappa} \bigg( \epsilon^{*}_{F} - \Gamma^{*} \bigg) 
                         \Theta \bigg(\epsilon^{*}_{F} - \Gamma^{*} \bigg)
                        \Bigg \{ 2\bigg[\tau^{*} + \frac{s_{m_{i}}}{4} \bigg]
                                 \bigg[F_{1} + \Big(1+ \frac{s_{m{i}}}{2} \Big) F_{2} \bigg]^{2}
                                   \nonumber \\
            &      &   - \Delta v \bigg[ \Big( \epsilon^{*}_{F} + \Gamma^{*}
                                  \Big)s_{m_{i}} +
                                        4 \lambda^{*} \Big(1+s_{m_{i}} \Big) \bigg]F_{1}F_{2} 
                             \nonumber \\ 
             &    &  +2\Delta v \bigg[\frac{1}{3} \Delta v 
                             \Big(\epsilon^{*^{2}}_{F} + \epsilon^{*}_{F} \Gamma^{*} + 
                                                 \Gamma^{*^{2}} \Big)+
                         \Delta v \Big(\epsilon^{*}_{F} + 
                     \Gamma^{*} \Big)\lambda^{*} -
                               \Delta v \tau^{*} + 
                                   \nonumber \\
           &        &  \Delta m^{*^{2}}  \Big( \epsilon^{*}_{F} + 
                                     \Gamma^{*} + 2 \lambda^{*} \Big) -
                        \Big( \Delta v + 2 \lambda^{*} \Big) 
                            \Big(1 + \frac{s_{m_{i}}}{2} \Big)^{2} \bigg]F^{2}_{2} 
                                        \nonumber \\
            &   & \bigg[ \frac{\tau^{*}}{\kappa} \Big( \frac{1}{3} \big( \epsilon^{*^{2}}_{F}
                                   + \epsilon^{*}_{F} \Gamma^{*} + \Gamma^{*^{2}} \big) + 
                                  \big(\epsilon^{*}_{F} + \Gamma^{*} \big) \lambda^{*} 
                                     \rho^{*}
                             + \lambda^{*^{2}} - \Delta m^{*^{2}} \big(1+ \rho^{*} \big)
                                             \Big) -
                                     \nonumber \\
            &   &     \Big(\tau^{*} + \big(1+s_{m_{i}} \big)^{2} \Big) \bigg] 
                          \bigg[ F^{2}_{1} + \tau F^{2}_{2} \bigg] \Bigg\} \, ,
\ea
where the $\Gamma^\ast$ function is given by 
\be
\Gamma^{*} \, = \, Max \Bigg\{ \bigg(  \epsilon^{*}_{F} - 2 \lambda^{*} \bigg) \, ; \, 
                        \bigg(-\lambda^{*} \rho^{*} + 
                    \kappa\sqrt{\rho^{*^{2}} + \frac{\Big(1+s_{m_{i}}\Big)^{2}}{\tau^{*}}} 
                    \bigg)  \Bigg\} \, .
\ee

% \linespread{0.5}
%\appendix
%\small
%\bibliographystyle{apsrev4-1}

%\bibliography{bibliography}
%\end{document}

%%%%%%%%%%%%%%%%%%%%%%%%%%%%%%%%%%%%%%%%%%%%%%%%%%%%%%%%%%%%%
\newpage
%\nonumsection{References}

\clearpage
\bibliography{biblio}

\begin{thebibliography}{36}
\expandafter\ifx\csname natexlab\endcsname\relax\def\natexlab#1{#1}\fi
\expandafter\ifx\csname bibnamefont\endcsname\relax
  \def\bibnamefont#1{#1}\fi
\expandafter\ifx\csname bibfnamefont\endcsname\relax
  \def\bibfnamefont#1{#1}\fi
\expandafter\ifx\csname citenamefont\endcsname\relax
  \def\citenamefont#1{#1}\fi
\expandafter\ifx\csname url\endcsname\relax
  \def\url#1{\texttt{#1}}\fi
\expandafter\ifx\csname urlprefix\endcsname\relax\def\urlprefix{URL }\fi
\providecommand{\bibinfo}[2]{#2}
\providecommand{\eprint}[2][]{\url{#2}}

\bibitem[{\citenamefont{Alvarez-Ruso et~al.}(2018)}]{Alvarez-Ruso:2017oui}
\bibinfo{author}{\bibfnamefont{L.}~\bibnamefont{Alvarez-Ruso}}
  \bibnamefont{et~al.}, \bibinfo{journal}{Prog. Part. Nucl. Phys.}
  \textbf{\bibinfo{volume}{100}}, \bibinfo{pages}{1} (\bibinfo{year}{2018}),
  \eprint{1706.03621}.

\bibitem[{\citenamefont{Katori and Martini}(2018)}]{Katori:2016yel}
\bibinfo{author}{\bibfnamefont{T.}~\bibnamefont{Katori}} \bibnamefont{and}
  \bibinfo{author}{\bibfnamefont{M.}~\bibnamefont{Martini}},
  \bibinfo{journal}{J. Phys.} \textbf{\bibinfo{volume}{G45}},
  \bibinfo{pages}{013001} (\bibinfo{year}{2018}), \eprint{1611.07770}.

\bibitem[{\citenamefont{Amaro et~al.}(2011)\citenamefont{Amaro, Barbaro,
  Caballero, Donnelly, and Williamson}}]{Amaro:2010sd}
\bibinfo{author}{\bibfnamefont{J.~E.} \bibnamefont{Amaro}},
  \bibinfo{author}{\bibfnamefont{M.~B.} \bibnamefont{Barbaro}},
  \bibinfo{author}{\bibfnamefont{J.~A.} \bibnamefont{Caballero}},
  \bibinfo{author}{\bibfnamefont{T.~W.} \bibnamefont{Donnelly}},
  \bibnamefont{and} \bibinfo{author}{\bibfnamefont{C.~F.}
  \bibnamefont{Williamson}}, \bibinfo{journal}{Phys. Lett.}
  \textbf{\bibinfo{volume}{B696}}, \bibinfo{pages}{151} (\bibinfo{year}{2011}),
  \eprint{1010.1708}.

\bibitem[{\citenamefont{Megias et~al.}(2016{\natexlab{a}})\citenamefont{Megias,
  Amaro, Barbaro, Caballero, and Donnelly}}]{Megias:2016lke}
\bibinfo{author}{\bibfnamefont{G.~D.} \bibnamefont{Megias}},
  \bibinfo{author}{\bibfnamefont{J.~E.} \bibnamefont{Amaro}},
  \bibinfo{author}{\bibfnamefont{M.~B.} \bibnamefont{Barbaro}},
  \bibinfo{author}{\bibfnamefont{J.~A.} \bibnamefont{Caballero}},
  \bibnamefont{and} \bibinfo{author}{\bibfnamefont{T.~W.}
  \bibnamefont{Donnelly}}, \bibinfo{journal}{Phys. Rev.}
  \textbf{\bibinfo{volume}{D94}}, \bibinfo{pages}{013012}
  (\bibinfo{year}{2016}{\natexlab{a}}), \eprint{1603.08396}.

\bibitem[{\citenamefont{Megias et~al.}(2017)\citenamefont{Megias, Barbaro,
  Caballero, Amaro, Donnelly, Ruiz~Simo, and Van~Orden}}]{Megias:2017cuh}
\bibinfo{author}{\bibfnamefont{G.~D.} \bibnamefont{Megias}},
  \bibinfo{author}{\bibfnamefont{M.~B.} \bibnamefont{Barbaro}},
  \bibinfo{author}{\bibfnamefont{J.~A.} \bibnamefont{Caballero}},
  \bibinfo{author}{\bibfnamefont{J.~E.} \bibnamefont{Amaro}},
  \bibinfo{author}{\bibfnamefont{T.~W.} \bibnamefont{Donnelly}},
  \bibinfo{author}{\bibfnamefont{I.}~\bibnamefont{Ruiz~Simo}},
  \bibnamefont{and} \bibinfo{author}{\bibfnamefont{J.~W.}
  \bibnamefont{Van~Orden}} (\bibinfo{year}{2017}), \eprint{1711.00771}.

\bibitem[{\citenamefont{Bodek et~al.}(2016)\citenamefont{Bodek, Christy, and
  Coopersmith}}]{Bodek:2016abf}
\bibinfo{author}{\bibfnamefont{A.}~\bibnamefont{Bodek}},
  \bibinfo{author}{\bibfnamefont{M.~E.} \bibnamefont{Christy}},
  \bibnamefont{and}
  \bibinfo{author}{\bibfnamefont{B.}~\bibnamefont{Coopersmith}},
  \bibinfo{journal}{Nucl. Part. Phys. Proc.}
  \textbf{\bibinfo{volume}{273-275}}, \bibinfo{pages}{1705}
  (\bibinfo{year}{2016}).

\bibitem[{\citenamefont{Bodek et~al.}(2015)\citenamefont{Bodek, Christy, and
  Coopersmith}}]{Bodek:2014jxa}
\bibinfo{author}{\bibfnamefont{A.}~\bibnamefont{Bodek}},
  \bibinfo{author}{\bibfnamefont{M.~E.} \bibnamefont{Christy}},
  \bibnamefont{and}
  \bibinfo{author}{\bibfnamefont{B.}~\bibnamefont{Coopersmith}},
  \bibinfo{journal}{AIP Conf. Proc.} \textbf{\bibinfo{volume}{1680}},
  \bibinfo{pages}{020003} (\bibinfo{year}{2015}), \eprint{1409.8545}.

\bibitem[{\citenamefont{Bodek et~al.}(2014)\citenamefont{Bodek, Christy, and
  Coopersmith}}]{Bodek:2014pka}
\bibinfo{author}{\bibfnamefont{A.}~\bibnamefont{Bodek}},
  \bibinfo{author}{\bibfnamefont{M.~E.} \bibnamefont{Christy}},
  \bibnamefont{and}
  \bibinfo{author}{\bibfnamefont{B.}~\bibnamefont{Coopersmith}},
  \bibinfo{journal}{Eur. Phys. J.} \textbf{\bibinfo{volume}{C74}},
  \bibinfo{pages}{3091} (\bibinfo{year}{2014}), \eprint{1405.0583}.

\bibitem[{\citenamefont{Kim and Wright}(2007)}]{Kim:2007yj}
\bibinfo{author}{\bibfnamefont{K.~S.} \bibnamefont{Kim}} \bibnamefont{and}
  \bibinfo{author}{\bibfnamefont{L.~E.} \bibnamefont{Wright}},
  \bibinfo{journal}{Phys. Rev.} \textbf{\bibinfo{volume}{C76}},
  \bibinfo{pages}{044613} (\bibinfo{year}{2007}), \eprint{0705.0049}.

\bibitem[{\citenamefont{Kim and Wright}(2003)}]{Kim:2003wy}
\bibinfo{author}{\bibfnamefont{K.~S.} \bibnamefont{Kim}} \bibnamefont{and}
  \bibinfo{author}{\bibfnamefont{L.~E.} \bibnamefont{Wright}},
  \bibinfo{journal}{Phys. Rev.} \textbf{\bibinfo{volume}{C67}},
  \bibinfo{pages}{054604} (\bibinfo{year}{2003}).

\bibitem[{\citenamefont{Giusti and Meucci}(2013)}]{Giusti:2013gsa}
\bibinfo{author}{\bibfnamefont{C.}~\bibnamefont{Giusti}} \bibnamefont{and}
  \bibinfo{author}{\bibfnamefont{A.}~\bibnamefont{Meucci}},
  \bibinfo{journal}{Nucl. Theor.} \textbf{\bibinfo{volume}{32}},
  \bibinfo{pages}{50} (\bibinfo{year}{2013}), \eprint{1309.4267}.

\bibitem[{\citenamefont{Meucci et~al.}(2013)\citenamefont{Meucci, Vorabbi,
  Giusti, Pacati, and Finelli}}]{Meucci:2013pua}
\bibinfo{author}{\bibfnamefont{A.}~\bibnamefont{Meucci}},
  \bibinfo{author}{\bibfnamefont{M.}~\bibnamefont{Vorabbi}},
  \bibinfo{author}{\bibfnamefont{C.}~\bibnamefont{Giusti}},
  \bibinfo{author}{\bibfnamefont{F.~D.} \bibnamefont{Pacati}},
  \bibnamefont{and} \bibinfo{author}{\bibfnamefont{P.}~\bibnamefont{Finelli}},
  \bibinfo{journal}{Phys. Rev.} \textbf{\bibinfo{volume}{C87}},
  \bibinfo{pages}{054620} (\bibinfo{year}{2013}), \eprint{1302.3390}.

\bibitem[{\citenamefont{Rocco et~al.}(2016)\citenamefont{Rocco, Lovato, and
  Benhar}}]{Rocco:2015cil}
\bibinfo{author}{\bibfnamefont{N.}~\bibnamefont{Rocco}},
  \bibinfo{author}{\bibfnamefont{A.}~\bibnamefont{Lovato}}, \bibnamefont{and}
  \bibinfo{author}{\bibfnamefont{O.}~\bibnamefont{Benhar}},
  \bibinfo{journal}{Phys. Rev. Lett.} \textbf{\bibinfo{volume}{116}},
  \bibinfo{pages}{192501} (\bibinfo{year}{2016}), \eprint{1512.07426}.

\bibitem[{\citenamefont{Benhar and Lovato}(2015)}]{Benhar:2015xga}
\bibinfo{author}{\bibfnamefont{O.}~\bibnamefont{Benhar}} \bibnamefont{and}
  \bibinfo{author}{\bibfnamefont{A.}~\bibnamefont{Lovato}},
  \bibinfo{journal}{Int. J. Mod. Phys.} \textbf{\bibinfo{volume}{E24}},
  \bibinfo{pages}{1530006} (\bibinfo{year}{2015}), \eprint{1506.05225}.

\bibitem[{\citenamefont{Benhar et~al.}(2010)\citenamefont{Benhar, Coletti, and
  Meloni}}]{Benhar:2010nx}
\bibinfo{author}{\bibfnamefont{O.}~\bibnamefont{Benhar}},
  \bibinfo{author}{\bibfnamefont{P.}~\bibnamefont{Coletti}}, \bibnamefont{and}
  \bibinfo{author}{\bibfnamefont{D.}~\bibnamefont{Meloni}},
  \bibinfo{journal}{Phys. Rev. Lett.} \textbf{\bibinfo{volume}{105}},
  \bibinfo{pages}{132301} (\bibinfo{year}{2010}), \eprint{1006.4783}.

\bibitem[{\citenamefont{Mosel and Gallmeister}(2019)}]{Mosel:2018qmv}
\bibinfo{author}{\bibfnamefont{U.}~\bibnamefont{Mosel}} \bibnamefont{and}
  \bibinfo{author}{\bibfnamefont{K.}~\bibnamefont{Gallmeister}},
  \bibinfo{journal}{Phys. Rev.} \textbf{\bibinfo{volume}{C99}},
  \bibinfo{pages}{064605} (\bibinfo{year}{2019}), \eprint{1811.10637}.

\bibitem[{\citenamefont{Caballero}(2006)}]{Caballero:2006wi}
\bibinfo{author}{\bibfnamefont{J.~A.} \bibnamefont{Caballero}},
  \bibinfo{journal}{Phys. Rev.} \textbf{\bibinfo{volume}{C74}},
  \bibinfo{pages}{015502} (\bibinfo{year}{2006}), \eprint{nucl-th/0604020}.

\bibitem[{\citenamefont{Maieron et~al.}(2002)\citenamefont{Maieron, Donnelly,
  and Sick}}]{Maieron:2001it}
\bibinfo{author}{\bibfnamefont{C.}~\bibnamefont{Maieron}},
  \bibinfo{author}{\bibfnamefont{T.~W.} \bibnamefont{Donnelly}},
  \bibnamefont{and} \bibinfo{author}{\bibfnamefont{I.}~\bibnamefont{Sick}},
  \bibinfo{journal}{Phys. Rev.} \textbf{\bibinfo{volume}{C65}},
  \bibinfo{pages}{025502} (\bibinfo{year}{2002}), \eprint{nucl-th/0109032}.

\bibitem[{\citenamefont{Megias et~al.}(2016{\natexlab{b}})\citenamefont{Megias,
  Amaro, Barbaro, Caballero, Donnelly, and Ruiz~Simo}}]{Megias:2016fjk}
\bibinfo{author}{\bibfnamefont{G.}~\bibnamefont{Megias}},
  \bibinfo{author}{\bibfnamefont{J.}~\bibnamefont{Amaro}},
  \bibinfo{author}{\bibfnamefont{M.}~\bibnamefont{Barbaro}},
  \bibinfo{author}{\bibfnamefont{J.}~\bibnamefont{Caballero}},
  \bibinfo{author}{\bibfnamefont{T.}~\bibnamefont{Donnelly}}, \bibnamefont{and}
  \bibinfo{author}{\bibfnamefont{I.}~\bibnamefont{Ruiz~Simo}},
  \bibinfo{journal}{Phys. Rev.} \textbf{\bibinfo{volume}{D94}},
  \bibinfo{pages}{093004} (\bibinfo{year}{2016}{\natexlab{b}}),
  \eprint{1607.08565}.

\bibitem[{\citenamefont{Gonz\'alez-Jim\'enez
  et~al.}(2014)\citenamefont{Gonz\'alez-Jim\'enez, Megias, Barbaro, Caballero,
  and Donnelly}}]{Gonzalez-Jimenez:2014eqa}
\bibinfo{author}{\bibfnamefont{R.}~\bibnamefont{Gonz\'alez-Jim\'enez}},
  \bibinfo{author}{\bibfnamefont{G.~D.} \bibnamefont{Megias}},
  \bibinfo{author}{\bibfnamefont{M.~B.} \bibnamefont{Barbaro}},
  \bibinfo{author}{\bibfnamefont{J.~A.} \bibnamefont{Caballero}},
  \bibnamefont{and} \bibinfo{author}{\bibfnamefont{T.~W.}
  \bibnamefont{Donnelly}}, \bibinfo{journal}{Phys. Rev.}
  \textbf{\bibinfo{volume}{C90}}, \bibinfo{pages}{035501}
  (\bibinfo{year}{2014}), \eprint{1407.8346}.

\bibitem[{\citenamefont{Amaro et~al.}(2005)\citenamefont{Amaro, Barbaro,
  Caballero, Donnelly, Molinari, and Sick}}]{Amaro:2004bs}
\bibinfo{author}{\bibfnamefont{J.~E.} \bibnamefont{Amaro}},
  \bibinfo{author}{\bibfnamefont{M.~B.} \bibnamefont{Barbaro}},
  \bibinfo{author}{\bibfnamefont{J.~A.} \bibnamefont{Caballero}},
  \bibinfo{author}{\bibfnamefont{T.~W.} \bibnamefont{Donnelly}},
  \bibinfo{author}{\bibfnamefont{A.}~\bibnamefont{Molinari}}, \bibnamefont{and}
  \bibinfo{author}{\bibfnamefont{I.}~\bibnamefont{Sick}},
  \bibinfo{journal}{Phys. Rev.} \textbf{\bibinfo{volume}{C71}},
  \bibinfo{pages}{015501} (\bibinfo{year}{2005}), \eprint{nucl-th/0409078}.

\bibitem[{\citenamefont{Amaro et~al.}(2012)\citenamefont{Amaro, Barbaro,
  Caballero, and Donnelly}}]{Amaro:2011aa}
\bibinfo{author}{\bibfnamefont{J.~E.} \bibnamefont{Amaro}},
  \bibinfo{author}{\bibfnamefont{M.~B.} \bibnamefont{Barbaro}},
  \bibinfo{author}{\bibfnamefont{J.~A.} \bibnamefont{Caballero}},
  \bibnamefont{and} \bibinfo{author}{\bibfnamefont{T.~W.}
  \bibnamefont{Donnelly}}, \bibinfo{journal}{Phys. Rev. Lett.}
  \textbf{\bibinfo{volume}{108}}, \bibinfo{pages}{152501}
  (\bibinfo{year}{2012}), \eprint{1112.2123}.

\bibitem[{\citenamefont{Caballero et~al.}(2005)\citenamefont{Caballero, Amaro,
  Barbaro, Donnelly, Maieron, and Udias}}]{Caballero:2005sj}
\bibinfo{author}{\bibfnamefont{J.~A.} \bibnamefont{Caballero}},
  \bibinfo{author}{\bibfnamefont{J.~E.} \bibnamefont{Amaro}},
  \bibinfo{author}{\bibfnamefont{M.~B.} \bibnamefont{Barbaro}},
  \bibinfo{author}{\bibfnamefont{T.~W.} \bibnamefont{Donnelly}},
  \bibinfo{author}{\bibfnamefont{C.}~\bibnamefont{Maieron}}, \bibnamefont{and}
  \bibinfo{author}{\bibfnamefont{J.~M.} \bibnamefont{Udias}},
  \bibinfo{journal}{Phys. Rev. Lett.} \textbf{\bibinfo{volume}{95}},
  \bibinfo{pages}{252502} (\bibinfo{year}{2005}), \eprint{nucl-th/0504040}.

\bibitem[{\citenamefont{Caballero et~al.}(2007)\citenamefont{Caballero, Amaro,
  Barbaro, Donnelly, and Udias}}]{Caballero:2007tz}
\bibinfo{author}{\bibfnamefont{J.~A.} \bibnamefont{Caballero}},
  \bibinfo{author}{\bibfnamefont{J.~E.} \bibnamefont{Amaro}},
  \bibinfo{author}{\bibfnamefont{M.~B.} \bibnamefont{Barbaro}},
  \bibinfo{author}{\bibfnamefont{T.~W.} \bibnamefont{Donnelly}},
  \bibnamefont{and} \bibinfo{author}{\bibfnamefont{J.~M.} \bibnamefont{Udias}},
  \bibinfo{journal}{Phys. Lett.} \textbf{\bibinfo{volume}{B653}},
  \bibinfo{pages}{366} (\bibinfo{year}{2007}), \eprint{0705.1429}.

\bibitem[{\citenamefont{Gonzalez-Jimenez
  et~al.}(2019)\citenamefont{Gonzalez-Jimenez, Barbaro, Caballero, Donnelly,
  N., Megias, Niewczas, Nikolakopoulos, and Udias}}]{Raul:2019}
\bibinfo{author}{\bibfnamefont{R.}~\bibnamefont{Gonzalez-Jimenez}},
  \bibinfo{author}{\bibfnamefont{M.}~\bibnamefont{Barbaro}},
  \bibinfo{author}{\bibfnamefont{J.}~\bibnamefont{Caballero}},
  \bibinfo{author}{\bibfnamefont{T.}~\bibnamefont{Donnelly}},
  \bibinfo{author}{\bibfnamefont{J.}~\bibnamefont{N.}},
  \bibinfo{author}{\bibfnamefont{G.}~\bibnamefont{Megias}},
  \bibinfo{author}{\bibfnamefont{K.}~\bibnamefont{Niewczas}},
  \bibinfo{author}{\bibfnamefont{A.}~\bibnamefont{Nikolakopoulos}},
  \bibnamefont{and} \bibinfo{author}{\bibfnamefont{J.}~\bibnamefont{Udias}},
  \bibinfo{journal}{submitted to Phys. Rev. C, arXiV:1909.07497}
  (\bibinfo{year}{2019}).

\bibitem[{\citenamefont{Horowitz and Piekarewicz}(1993)}]{Horowitz:1993nb}
\bibinfo{author}{\bibfnamefont{C.~J.} \bibnamefont{Horowitz}} \bibnamefont{and}
  \bibinfo{author}{\bibfnamefont{J.}~\bibnamefont{Piekarewicz}},
  \bibinfo{journal}{Phys. Rev.} \textbf{\bibinfo{volume}{C47}},
  \bibinfo{pages}{2924} (\bibinfo{year}{1993}), \eprint{nucl-th/9302004}.

\bibitem[{\citenamefont{Kim et~al.}(1995)\citenamefont{Kim, Horowitz, and
  Frank}}]{Kim:1994nr}
\bibinfo{author}{\bibfnamefont{H.-c.} \bibnamefont{Kim}},
  \bibinfo{author}{\bibfnamefont{C.~J.} \bibnamefont{Horowitz}},
  \bibnamefont{and} \bibinfo{author}{\bibfnamefont{M.~R.} \bibnamefont{Frank}},
  \bibinfo{journal}{Phys. Rev.} \textbf{\bibinfo{volume}{C51}},
  \bibinfo{pages}{792} (\bibinfo{year}{1995}), \eprint{nucl-th/9410015}.

\bibitem[{\citenamefont{Donnelly and Raskin}(1986)}]{Donnelly:1985ry}
\bibinfo{author}{\bibfnamefont{T.~W.} \bibnamefont{Donnelly}} \bibnamefont{and}
  \bibinfo{author}{\bibfnamefont{A.~S.} \bibnamefont{Raskin}},
  \bibinfo{journal}{Annals Phys.} \textbf{\bibinfo{volume}{169}},
  \bibinfo{pages}{247} (\bibinfo{year}{1986}).

\bibitem[{\citenamefont{Donnelly and Sick}(1999)}]{Donnelly:1998xg}
\bibinfo{author}{\bibfnamefont{T.~W.} \bibnamefont{Donnelly}} \bibnamefont{and}
  \bibinfo{author}{\bibfnamefont{I.}~\bibnamefont{Sick}},
  \bibinfo{journal}{Phys. Rev. Lett.} \textbf{\bibinfo{volume}{82}},
  \bibinfo{pages}{3212} (\bibinfo{year}{1999}), \eprint{nucl-th/9809063}.

\bibitem[{\citenamefont{Amaro et~al.}(2002)\citenamefont{Amaro, Barbaro,
  Caballero, Donnelly, and Molinari}}]{Amaro:2002mj}
\bibinfo{author}{\bibfnamefont{J.~E.} \bibnamefont{Amaro}},
  \bibinfo{author}{\bibfnamefont{M.~B.} \bibnamefont{Barbaro}},
  \bibinfo{author}{\bibfnamefont{J.~A.} \bibnamefont{Caballero}},
  \bibinfo{author}{\bibfnamefont{T.~W.} \bibnamefont{Donnelly}},
  \bibnamefont{and} \bibinfo{author}{\bibfnamefont{A.}~\bibnamefont{Molinari}},
  \bibinfo{journal}{Phys. Rept.} \textbf{\bibinfo{volume}{368}},
  \bibinfo{pages}{317} (\bibinfo{year}{2002}), \eprint{nucl-th/0204001}.

\bibitem[{\citenamefont{De~Forest}(1983)}]{DeForest:1983ahx}
\bibinfo{author}{\bibfnamefont{T.}~\bibnamefont{De~Forest}},
  \bibinfo{journal}{Nucl. Phys.} \textbf{\bibinfo{volume}{A392}},
  \bibinfo{pages}{232} (\bibinfo{year}{1983}).

\bibitem[{\citenamefont{Galster et~al.}(1971)\citenamefont{Galster, Klein,
  Moritz, Schmidt, Wegener, and Bleckwenn}}]{Galster:1971kv}
\bibinfo{author}{\bibfnamefont{S.}~\bibnamefont{Galster}},
  \bibinfo{author}{\bibfnamefont{H.}~\bibnamefont{Klein}},
  \bibinfo{author}{\bibfnamefont{J.}~\bibnamefont{Moritz}},
  \bibinfo{author}{\bibfnamefont{K.~H.} \bibnamefont{Schmidt}},
  \bibinfo{author}{\bibfnamefont{D.}~\bibnamefont{Wegener}}, \bibnamefont{and}
  \bibinfo{author}{\bibfnamefont{J.}~\bibnamefont{Bleckwenn}},
  \bibinfo{journal}{Nucl. Phys.} \textbf{\bibinfo{volume}{B32}},
  \bibinfo{pages}{221} (\bibinfo{year}{1971}).

\bibitem[{\citenamefont{Alberico et~al.}(1988)\citenamefont{Alberico, Molinari,
  Donnelly, Kronenberg, and Van~Orden}}]{Alberico:1988bv}
\bibinfo{author}{\bibfnamefont{W.~M.} \bibnamefont{Alberico}},
  \bibinfo{author}{\bibfnamefont{A.}~\bibnamefont{Molinari}},
  \bibinfo{author}{\bibfnamefont{T.~W.} \bibnamefont{Donnelly}},
  \bibinfo{author}{\bibfnamefont{E.~L.} \bibnamefont{Kronenberg}},
  \bibnamefont{and} \bibinfo{author}{\bibfnamefont{J.~W.}
  \bibnamefont{Van~Orden}}, \bibinfo{journal}{Phys. Rev.}
  \textbf{\bibinfo{volume}{C38}}, \bibinfo{pages}{1801} (\bibinfo{year}{1988}).

\bibitem[{\citenamefont{Alberico et~al.}(1990)\citenamefont{Alberico, Donnelly,
  and Molinari}}]{Alberico:1989aja}
\bibinfo{author}{\bibfnamefont{W.~M.} \bibnamefont{Alberico}},
  \bibinfo{author}{\bibfnamefont{T.~W.} \bibnamefont{Donnelly}},
  \bibnamefont{and} \bibinfo{author}{\bibfnamefont{A.}~\bibnamefont{Molinari}},
  \bibinfo{journal}{Nucl. Phys.} \textbf{\bibinfo{volume}{A512}},
  \bibinfo{pages}{541} (\bibinfo{year}{1990}).

\bibitem[{\citenamefont{Benhar et~al.}(2008)\citenamefont{Benhar, Day, and
  Sick}}]{Benhar:2006wy}
\bibinfo{author}{\bibfnamefont{O.}~\bibnamefont{Benhar}},
  \bibinfo{author}{\bibfnamefont{D.}~\bibnamefont{Day}}, \bibnamefont{and}
  \bibinfo{author}{\bibfnamefont{I.}~\bibnamefont{Sick}},
  \bibinfo{journal}{Rev. Mod. Phys.} \textbf{\bibinfo{volume}{80}},
  \bibinfo{pages}{189} (\bibinfo{year}{2008}), \eprint{nucl-ex/0603029}.

\bibitem[{\citenamefont{Benhar et~al.}(2006)\citenamefont{Benhar, Day, and
  Sick}}]{Benhar:2006er}
\bibinfo{author}{\bibfnamefont{O.}~\bibnamefont{Benhar}},
  \bibinfo{author}{\bibfnamefont{D.}~\bibnamefont{Day}}, \bibnamefont{and}
  \bibinfo{author}{\bibfnamefont{I.}~\bibnamefont{Sick}}
  (\bibinfo{year}{2006}), \eprint{nucl-ex/0603032}.

\end{thebibliography}

\end{document}